\documentclass[superscriptaddress,aps,preprintnumbers,amsmath,showpacs,amssymb,prd,nofootinbib,reprint]{revtex4-1}
\pdfoutput=1
\usepackage{booktabs} 
\usepackage{array} 
\usepackage[table,xcdraw]{xcolor} 
\usepackage{amsmath,amssymb}
\usepackage{graphicx}
\usepackage{fancyhdr}
\usepackage{bm, color}
\usepackage{slashed,amsthm,amsfonts,empheq}
\usepackage[caption=false]{subfig}
\usepackage{hyperref}
\hypersetup{hidelinks}
\usepackage{booktabs}
\usepackage{multirow}
\usepackage[left=2cm,right=2cm,top=2cm,bottom=2cm,includefoot,a4paper]{geometry}
\usepackage{tikz}
\usepackage{placeins}
\usetikzlibrary{arrows.meta}
\usepackage{amsmath}

\newcommand{\Slash}[1]{{\ooalign{\hfil#1\hfil\crcr\raise.167ex\hbox{/}}}}

\newcommand{\beq}{\begin{equation}}  \newcommand{\eeq}{\end{equation}}
\newcommand{\bef}{\begin{figure}}  \newcommand{\eef}{\end{figure}}
\newcommand{\bec}{\begin{center}}  \newcommand{\eec}{\end{center}}
\newcommand{\non}{\nonumber}  
\newcommand{\dd}{\mathrm{d}}

\newcommand{\laq}[1]{\label{eq:#1}}  
\newcommand{\Eq}[1]{Eq.~(\ref{eq:#1})}

\newcommand{\Sec}[1]{Sec.~\ref{chap:#1}}
\newcommand{\ab}[1]{\left|{#1}\right|}
\newcommand{\vev}[1]{\left\langle {#1} \right\rangle}

\newcommand{\lac}[1]{\label{chap:#1}}
\newcommand{\SU}[1]{{\rm SU{#1} } }

\def\({\left(}
\def\){\right)}
\def\dt{\frac{d}{dt}}

\def\O{\mathcal{O}}
\def\U{\mathop{\rm U}}

\newcommand{\EV}{\,{\rm eV}}
\newcommand{\KEV}{\,{\rm keV}}
\newcommand{\MEV}{\,{\rm MeV}}
\newcommand{\GEV}{\,{\rm GeV}}
\newcommand{\TEV}{\,{\rm TeV}}
\def\o{\over}
\def\a{\alpha}
\def\b{\beta}

\def\f{\phi}
\def\g{\gamma}
\def\h{\theta}
\def\k{\kappa}

\def\m{\mu}

\def\q{\partial}
\def\r{\rho}

\def\y{\eta}

\def\D{\Delta}
\def\G{\Gamma}

\def\me{\mathrm{e}}
\def\ol{\overline}

\def\*{\dagger}

\begin{document}

\title{
{$\mu$DM: a new mechanism {for} the baryon--dark matter coincidence}
}

\author{Wen Yin}
\affiliation{Department of Physics, Tokyo Metropolitan University, Minami-Osawa, Hachioji-shi, Tokyo 192-0397, Japan}

\begin{abstract}
{For a relativistically decoupled thermal relic, the dark-matter
energy-to-entropy ratio scales as $\rho_{\rm DM}/s\sim 10^{-3}m_{\rm DM}$.
{This is the familiar hot-dark-matter abundance relation.}
 Immediately after weak-sphaleron freeze-out, the baryon asymmetry can be parametrized as
$\rho_{\rm baryon}/s\sim 10^{-4}|\mu_B|$.
Here $\mu_B$ is the baryon-number chemical
potential.
We propose a new class of models, called
chemical-potential-matched dark matter ($\mu$DM), in which
$|\mu_B|\sim m_{\rm DM}$, thereby providing a new route to the baryon--dark
matter coincidence.
As a concrete example of $\mu$DM, we consider an Affleck--Dine field
whose excitations constitute dark matter.  We also briefly discuss spontaneous
electroweak baryogenesis associated with axion-like-particle walls.  In the
absence of entropy dilution, this coincidence mechanism generically points
to dark matter in the eV--keV mass range.  Its momentum distribution
can nevertheless be cold if thermal production is dominated by Bose-enhanced
stimulated emission.  
{Alternatively, entropy dilution after both the dark-matter and baryon
yields are fixed allows masses up to the MeV scale.}}

\end{abstract}

\maketitle

\section{Introduction}

{Dark matter and baryons are both essential ingredients of our
Universe.}
{Dark matter and baryons have very different microscopic properties, yet
their measured cosmological abundances exhibit the numerical coincidence}~\cite{Planck:2018vyg}
\begin{equation}
{{\Omega_{\rm DM}\o\Omega_b}\simeq 5.36\pm0.07.}
 \laq{measuredratio}
\end{equation}
{This coincidence is mysterious because their standard production
mechanisms are usually very different.}

{One broad class of historical explanations} relates a dark-sector number asymmetry to
the baryon asymmetry {at the level of particle number rather than energy density}.  In asymmetric-dark-matter scenarios, this relation
typically points to a dark-matter mass near the nucleon mass if the two number
asymmetries are comparable~\cite{Nussinov:1985xr,Kaplan:1991ah,Kaplan:2009ag,Zurek:2013wia,Kitano:2004sv}.
The coincidence can also arise from nonthermal dark matter produced in the
decay of $Q$-balls formed by an Affleck--Dine condensate
~\cite{Fujii:2002aj,Roszkowski:2006kw}, or when the visible- and dark-sector
asymmetries are produced together by Hawking radiation~\cite{Hook:2014mla}. 
{In such mechanisms, further relating the dark-matter mass to the baryon mass can
account for the coincidence}~{\cite{Foot:2003jt, Newstead:2014jva,Farina:2015uea,Lonsdale:2018xwd,Ibe:2019ena, Murgui:2021eqf}}. 
More recently, a dynamical mechanism in
which a scalar adjusts the baryon and dark-matter masses until their energy
densities become comparable was proposed in
Refs.~\cite{Brzeminski:2023wza,Banerjee:2024xhn}.  
{Here we instead propose a new class of dark matter in which the two
energy densities are directly related.}

{Immediately after weak-sphaleron freeze-out, baryon number is
effectively conserved.}
Let $\mu_B$ denote the corresponding baryon-number chemical
potential.
{If it is dynamically related to the dark-matter mass as}
\beq \boxed{\laq{murela}|\mu_B| \sim m_{X}}\eeq  
where $m_X$ is the dark-matter mass, the baryon--dark matter
coincidence follows for a thermal-scale relic abundance.  We call this new class of dark matter
\textit{chemical-potential-matched dark matter} ($\mu$DM).

{The mechanism rests on two ingredients.  First, a relativistically
decoupled thermal relic has $n_X\sim T^3$ and hence}
\beq\rho_{X} \sim m_{X} T^3.\eeq
If such a relic constitutes all of dark matter, its free streaming
erases excessive small-scale structure; this is the familiar problem with
hot dark matter~\cite{Davis:1985rj}.  A light boson can, however, be
produced from a hot plasma with a cold momentum distribution through
Bose-enhanced stimulated emission~\cite{Yin:2023jjj,Sakurai:2024apm} (see also \cite{Moroi:2020has,Moroi:2020bkq}).
{Alternatively, late entropy production can dilute a relativistically
decoupled relic and reduce its free-streaming velocity}~{\cite{Patwardhan:2015kga}}. 
Unlike ordinary hot dark matter, {a relic produced by stimulated emission
or cooled by entropy dilution} can therefore be compatible
with structure formation.

Second, the frozen baryon asymmetry is related to $\mu_B$ by
\begin{equation}
|n_B(T_{\rm sph})|\sim
 |\mu_B(T_{\rm sph})|T_{\rm sph}^2,
 \laq{nbT2intro}
\end{equation}
This parametric relation does not assume a particular baryogenesis
mechanism.  The frozen density represented by $\mu_B$ may arise, for example,
from a pre-existing $B-L$ asymmetry or from an external time-dependent bias
acting before freeze-out
~\cite{Khlebnikov:1988sr,Harvey:1990qw,Burnier:2005hp}.
Spontaneous
baryogenesis is one concrete realization of the latter
~\cite{Cohen:1987vi}.  The baryon
rest-mass density is $\rho_B=m_N|n_B|$.  If Eq.~\eqref{eq:murela} holds, the
dark-matter mass cancels from the present-day ratio of energy densities.  The
cancellation also persists under a common entropy dilution after both the
dark-matter and baryon yields have been fixed.

{The purpose of this paper is to show that Eq.~\eqref{eq:murela} can arise
naturally.  One possibility is spontaneous baryogenesis at an electroweak
phase boundary induced by an axion-like-particle (ALP) domain wall
~\cite{2604.20762}.  In that case, the chemical-potential bias is controlled
by the wall width.  A separate mechanism is then required to avoid
overproducing ALPs when the walls disappear.  One possible
model-building solution is stimulated emission of the ALP through the inverse
of the process discussed in this paper and in Refs.~\cite{Yin:2023jjj,Sakurai:2024apm}, since
the steady state is an attractor.}
Our main example is an Affleck--Dine field moving on an approximately
circular trajectory.  For a quadratic potential, its angular velocity is
$\dot\theta\simeq m_R$.  The rotating phase then generates a bias of
the same order as the mass scale of the Affleck--Dine field; the resulting frozen
baryon density determines $\mu_B$.
{The Affleck--Dine field therefore realizes $\mu$DM.}  This construction combines
the Affleck--Dine mechanism~\cite{Affleck:1984fy} with spontaneous
baryogenesis~\cite{Cohen:1987vi,Cohen:1988kt, Davoudiasl:2004gf} or spontaneous leptogenesis~\cite{Li:2001st,Kusenko:2014lra,Kusenko:2014uta,Ibe:2015nfa} (see also Ref.~\cite{Fukugita:1986hr}){, as previously explored
for flat directions in Refs.~\cite{Takahashi:2003db,Chiba:2003vp}}.

\section{A proposal for $\mu$DM}
\lac{general}

\subsection{Thermal-scale relic
abundance}
{We first consider a dark-matter relic whose abundance is generated from
a thermal plasma and whose final number density is of thermal order. 
{This condition does not require
a thermal final-state momentum distribution, as discussed below.}

  Let $T_p$
denote the cosmic temperature at which the final comoving abundance of a bosonic
dark-matter particle $X$ is fixed, and define $\kappa_X$ by}
\begin{equation}
 n_X(T_p)=\k_XT_p^3,
 \qquad
 Y_X={n_X\o s}
 ={45\k_X\o2\pi^2g_{*s}(T_p)}.
 \laq{YX}
\end{equation}
The present energy density is
\begin{equation}
 \r_{X,0}=m_XY_Xs_0.
 \laq{rhoX0}
\end{equation}
{With ordinary relativistic decoupling, the relic retains a
thermal spectrum and is the familiar hot dark matter~{\cite{Bond:1980ha,Davis:1985rj}}. In particular, a single degree of freedom produced before the electroweak crossover predicts $\k_X\sim 0.1$.   If this initially hot
relic is subsequently cooled by entropy production, its free-streaming velocity
is reduced and it can instead behave as warm dark matter{~\cite{Scherrer:1984fd,Bode:2000gq}}.
{The entropy-dilution factor is}
$
 \Delta_S\equiv {S_{\rm after}/ S_{\rm before}},
 \rho_{X,0}\ \longrightarrow\ {\rho_{X,0}/\Delta_S},
 {T_X/ T}\ \longrightarrow\
 \Delta_S^{-1/3}{T_X/T},
$
{where $S=sa^3$ is the comoving entropy of the visible plasma.  The last relation means that $X$ is cooled relative to the visible plasma.}
Let $T_M$ denote the temperature at the onset of matter domination and $T_{\rm RH}$ the reheating temperature.  For a matter-dominated era beginning after the electroweak crossover, $\D_S\simeq T_M/T_{\rm RH}$ up to order-one factors.  Requiring $T_{\rm RH}\gtrsim4\MEV$~\cite{deSalas:2015glj} therefore gives $\D_S\lesssim\text{a few}\times10^4$.

  Alternatively, Bose-enhanced production from a thermal plasma can concentrate the population in infrared modes
and {produce} a cold distribution while retaining the thermal-scale abundance in
Eq.~\eqref{eq:YX} characteristic of hot dark matter.  We call {this} possibility cold ``hot dark matter''~{\cite{Yin:2023jjj,Sakurai:2024apm}}.
Its stimulated-emission realization is summarized in
Appendix~\ref{app:coldhot}.}

{Consequently, the abundance fixes the mass to}
\begin{align}
 m_X&={\Delta_S}\,{\Omega_X\rho_c\o s_0}
 {2\pi^2g_{*s}(T_p)\o45\k_X}
 \nonumber\\
 &\simeq 230\EV\,{\Delta_S}
 \left({\Omega_Xh^2\o0.12}\right)
 \left({0.09\o\k_X}\right)
 \left({g_{*s}(T_p)\o110}\right).
 \laq{massprediction}
\end{align}
{We use $\k_X\simeq0.09$ as a benchmark, although it can range depending on the production environment; see
Appendix~\ref{app:coldhot}.}

Unless stated otherwise, we remain agnostic about the detailed production
process and parametrize the final abundance by $\kappa_X$.

\subsection{\texorpdfstring{Baryon asymmetry and the baryon-number
chemical potential}{Baryon asymmetry and the baryon-number chemical potential}}
Immediately after weak-sphaleron freeze-out, baryon number is
effectively conserved.  We characterize the frozen asymmetry parametrically
by the baryon-number chemical potential $\mu_B$,
\begin{equation}
 |n_B(T_{\rm sph})|\sim
 |\mu_B(T_{\rm sph})|T_{\rm sph}^2.
 \laq{nBeq}
\end{equation}
This relation does not specify how the asymmetry was generated:
the frozen baryon density may arise, for example, from a pre-existing $B-L$
asymmetry or from a time-dependent bias acting before freeze-out
~\cite{Khlebnikov:1988sr,Harvey:1990qw,Burnier:2005hp}.

Spontaneous baryogenesis provides one concrete realization
~\cite{Cohen:1987vi,Cohen:1988kt,Domcke:2020kcp}.  Let the scalar variable
$\theta$ couple to a baryon-number current,
\begin{equation}
 {\cal L}_{\rm SB}=-c\q_\mu\h J^\mu.
 \laq{SBcoupling}
\end{equation}
{In the homogeneous plasma frame,}
\begin{equation}
 \mu_{\rm bias}=c\dot\h.
 \laq{chemicalpotential}
\end{equation}
In this realization, the bias in Eq.~\eqref{eq:chemicalpotential}
generates the frozen baryon density and thereby determines $\mu_B$.

In the absence of subsequent baryon-number violation or production,
the comoving baryon number is conserved after sphaleron decoupling.  We
use the sphaleron-decoupling temperature~\cite{DOnofrio:2014rug},
\begin{equation}
{{T_{\rm sph}}\simeq131.7\GEV.}
 \laq{sphalerontemperature}
\end{equation}
{The baryon yield and its present rest-mass density are then}
\begin{equation}
Y_B\sim{45\over2\pi^2g_{*s}({T_{\rm sph}})}
 {{|\mu_B|}\over {T_{\rm sph}}},
 \qquad
 \rho_{B,0}=m_Ns_0Y_B.
 \laq{rhoB0}
\end{equation}
Entropy production after both the baryon and dark-matter yields are
fixed gives $Y_B\to Y_B/\D_S$ and $\rho_{B,0}\to\rho_{B,0}/\D_S$, the same
scaling as for dark matter.

\subsection{The baryon--dark matter ratio and the proposal for $\mu$DM}

Combining the thermal-scale relic abundance with the general
chemical-potential relation gives
\begin{equation}
 {\Omega_X\o\Omega_b}
 \sim\k_X
 {g_{*s}({T_{\rm sph}})\o g_{*s}(T_p)}
 {m_X\o|\mu_B|}{{T_{\rm sph}}\o m_N}.
 \laq{masterrelation}
\end{equation}
{Here $g_{*s}$ counts the relativistic degrees of freedom in the entropy
density.} This relation continues to hold under entropy production
after both yields are fixed because the common dilution factor cancels.
{Writing}
\begin{equation}
 |\mu_B|=c_\mu m_X,
 \laq{mumass}
\end{equation}
{the observed ratio requires}
\begin{multline}
 c_\mu\sim1.97{\k_X\o 0.1}
 {g_{*s}({T_{\rm sph}})\o g_{*s}(T_p)}
 \\
 \times\left({{T_{\rm sph}}\o100\GEV}\right)
 \left({0.94\GEV\o m_N}\right)
 \left({5.4\o\Omega_X/\Omega_b}\right).
 \laq{cmurequired}
\end{multline}
{Thus a scenario with
\Eq{murela} and a thermal-scale relic abundance}
naturally explains the coincidence. {We therefore propose a new class of dark-matter models: \textit{chemical-potential-matched dark matter} ($\mu$DM).}

We now outline two spontaneous-baryogenesis realizations of this more
general relation.

\begin{itemize}
\item {An ALP domain wall may induce an electroweak phase boundary and
source spontaneous baryogenesis~\cite{2604.20762}.  When
$m_X$ denotes the ALP mass in the range of interest, the resulting
chemical potential can obey $|\mu_B|\sim\gamma\beta m_X$, where $\beta$ and
$\gamma$ are the wall velocity and Lorentz factor, respectively.  Thus,
$\beta\gamma=\O(1)$ can realize the required relation naturally.  The associated ALP constitutes dark matter.  A potential difficulty is that collapse of the walls may
overproduce ALPs.  Efficient dissipation of the ALP condensate could alleviate
this problem.  One possible source of such efficient dissipation is
a backreaction process analogous to stimulated emission.  Alternatively, {an ALP-driven first-order electroweak phase transition may provide the relevant wall profile~\cite{Jeong:2018ucz}}, in which case the tension may not need to be too large.  We leave these possibilities for
future work.}
\item {An Affleck--Dine field may source spontaneous baryogenesis while
its excitations constitute dark matter.  This is our main focus.}
\end{itemize}

\section{Affleck--Dine field dark matter}
\lac{AD}
{For simplicity, we assume a radiation-dominated Universe throughout this section.}
\subsection{The relation $\dot\h\simeq m_R$}

Let
\begin{equation}
X= P={R\o\sqrt2}\me^{i\h}
 \laq{polarfield}
\end{equation}
be a complex scalar with an approximate $\U(1)$ symmetry.  We assume that its
potential is approximately quadratic at the relevant field value,
\begin{equation}
 V(P)\simeq m_R^2|P|^2,
 \laq{ADpotential}
\end{equation}
with $m_R^2>0$.
After a small $\U(1)$-breaking interaction gives an
initial torque, the charge is approximately conserved as in the usual
Affleck--Dine mechanism~\cite{Affleck:1984fy}:
\begin{equation}
 {\dd\over\dd t}\left[a^3R^2\dot\h\right]=0.
 \laq{ADcharge}
\end{equation}
{Here $a$ is the scale factor.}
The radial equation is
\begin{equation}
 \ddot R+3H\dot R-R\dot\h^2+m_R^2R=0.
 \laq{radialeom}
\end{equation}
{For a circular trajectory, or an elliptic trajectory with order-one
ellipticity, radial force balance gives the approximate solution}
\begin{equation}
 \dot\h\simeq m_R.
 \laq{ADrotation}
\end{equation}
Dissipation that respects the global $\U(1)$ symmetry decreases the
energy without violating the associated Noether charge.  Since the circular
trajectory minimizes the energy at fixed charge, it is an attractor.

Then, from Eq.~\eqref{eq:chemicalpotential}, we obtain
\begin{equation}
\mu_{\rm bias}\simeq cm_R.
 \laq{ADmu}
\end{equation}
This bias generates the frozen baryon density and thereby determines
$\mu_B$.
{The rotation can persist through sphaleron decoupling.}

The charge density and rotational energy density are
\begin{equation}
{n_P=R^2\dot\h,
 \qquad
 \r_{\rm rot}=m_R|n_P|=m_R^2R^2}
 \laq{ADdensity}
\end{equation}
{Equivalently, the two Cartesian components of $P$ undergo the ordinary misalignment
mechanism with a relative phase of $\pi/2$~\cite{Preskill:1982cy,Abbott:1982af,Dine:1982ah}.}

At sphaleron freeze-out, the plasma contribution to the conserved
charge is $cn_B$.  Using Eq.~\eqref{eq:nBeq}, its backreaction on the rotation
is small if
\begin{equation}
\begin{aligned}
 R^2m_R&\gg |cn_B|,\\
 R&\gg\left({|cn_B|\over m_R}\right)^{1/2}.
\end{aligned}
 \laq{ADbackreaction}
\end{equation}

Let $\k_P$ denote the thermal-scale abundance coefficient of the
$P$ excitations, including both radial and phase (axion-like) modes, defined
analogously to $\k_X$ in Eq.~\eqref{eq:YX}.
  {The rotating background is
subdominant to these excitations if}
\begin{equation}
 \begin{aligned}
 \r_{\rm rot}&\ll\r_P(T)
 =\k_Pm_R{g_{*s}(T)\o g_{*s}(T_p)}T^3,\\
 R&\ll\left[
 \k_P{g_{*s}(T)\o g_{*s}(T_p)}{T^3\o m_R}
 \right]^{1/2}.
 \end{aligned}
 \laq{rotationwindow}
\end{equation}
{For $\Delta_S=1$, Eq.~\eqref{eq:massprediction} gives $m_R=230\EV$ for
$\k_P=0.09$.  At $T=T_{\rm sph}$, the broad window can be written as}
\begin{equation}
\begin{aligned}
R&\gg131.7\GEV
\left({|c\mu_B|\over m_R}\right)^{1/2}
\left({T_{\rm sph}\over131.7\GEV}\right),\\
R&\ll9.5\times10^5\GEV
\left({\k_P\over0.09}\right)^{1/2}\\[-2pt]
&\quad\times
\left({g_{*s}(T_{\rm sph})\over g_{*s}(T_p)}\right)^{1/2}\\[-2pt]
&\quad\times
\left({230\EV\over m_R}\right)^{1/2}
\left({T_{\rm sph}\over131.7\GEV}\right)^{3/2}.
\end{aligned}
 \laq{rotationwindownumerical}
\end{equation}
Thus the same complex field can generate the bias that fixes the
frozen baryon density and hence $\mu_B$ while
the excitations of the thermal-scale relic dominate the dark-matter abundance.

\subsection{$\nu$ realization of $\mu$DM}
{Directly coupling $P$ to Standard Model particles generally induces a
strong thermal correction to its potential, as discussed below.}\footnote{Coupling to Standard Model particles is viable in the warm-dark-matter scenario mentioned in this paper, with entropy dilution.  Another simple but exotic possibility is to change the value of $P$ after sphaleron decoupling.  A coupling $AP|H|^2$ can shift the minimum of $P$ substantially during the Higgs phase transition without changing the mass of $P$.  This can also render the one-flavor toy model viable.}
This motivates couplings to another sector, such as
right-handed neutrinos that generate neutrino masses through the seesaw
mechanism~{\cite{Minkowski:1977sc,Yanagida:1979as,Gell-Mann:1979vob,Mohapatra:1979ia}}.

\paragraph{A toy model for illustration}
To illustrate this, let us first consider the single-flavor case.
We couple the rotating background to a right-handed neutrino $\nu_{R,1}$
that carries a $\U(1)$ charge equal to $-n/2$ times that of $P$;
this symmetry may be identified with lepton number:
\begin{multline}
 {\cal L}\supset-m_R^2|P|^2-y_{\a\nu}\ol{L_\a}\widetilde H\nu_{R,1}
 -{1\over2}y_PP^n\overline{\nu_{R,1}^{\,c}}\nu_{R,1}
+{\rm h.c.}\\
 \laq{lagrangian}
\end{multline}
A renormalizable realization with $n>1$ requires additional fields.
For simplicity, we integrate them out and work with the resulting effective
Lagrangian.
A chiral field redefinition that removes the phase of $P^n$ produces a
derivative coupling of the type introduced for spontaneous
leptogenesis in Refs.~\cite{Li:2001st,Kusenko:2014lra,Kusenko:2014uta,Ibe:2015nfa} and fixes the rotation-induced bias to
\begin{equation}
 \m_{\nu_{R,1}}^{\rm bias}={n\dot\h\over2}\simeq{n m_R\over2}.
 \laq{RHNChemicalPotential}
\end{equation}

In this one-flavor model, we parametrize the slowest transfer rate as
$
 \G_{\rm tr}^{(1)}(T)=\g_N\sum_\a
 {m_M(T)^2\o T^2}|y_{\a\nu}|^2T, ~\g_N\sim10^{-2}.
$ 
Here
$
 m_M(T)\equiv |y_P|\left({R(T)\o\sqrt2}\right)^n,
$
and the factor $m_M(T)^2/T^2$ represents chirality-flip suppression~\cite{Akhmedov:1998qx}.  The conditions relevant for
baryogenesis are
$
 \G_{\rm tr}^{(1)}(T_{\rm sph})\gtrsim H(T_{\rm sph}),\qquad T_{\rm sph} \gg m_M(T_{\rm sph}).
$
These conditions can be satisfied simultaneously.

While electroweak sphalerons are active, the plasma follows the biased
equilibrium.  Assuming that all three active-lepton flavors are equilibrated,
the chemical-potential constraints in
Appendix~\ref{app:twoRHNtransport} give
\begin{equation}
 \ab{n_B(T_{\rm sph})}={7\over33}n\ab{\dot\h}T_{\rm sph}^2
 ={7\over33}n m_RT_{\rm sph}^2
 \laq{RHNBaryonDensity}
\end{equation}
After sphaleron decoupling, the comoving baryon number is conserved,
even if the sterile-neutrino abundance and mass subsequently change.  Because
$m_M(T_{\rm sph})<T_{\rm sph}$ is required for the mechanism to operate, the
later decoupling or decay of the sterile sector into Standard Model particles
does not wash out the baryon asymmetry after sphaleron freeze-out, although it
may generate an additional lepton asymmetry.

This minimal model is not phenomenologically viable.  For the
mechanism to work, the thermal contribution to the $P$ potential must remain
subdominant.  The leading high-temperature term is
$\Delta V_T=m_M(R)^2T^2/24$; hence comparing its radial force with
$m_R^2R$ requires
\begin{equation}
 {n\over12}m_M(T)^2T^2\ll m_R^2R(T)^2.
 \laq{thermalmassbound}
\end{equation}
At the same time, requiring the coherent rotation to remain subdominant
to the measured dark-matter abundance gives
 \begin{align}
 m_M(T)&<1.4\MEV
\left({\k_P\o0.09}{g_{*s}(T)\o g_{*s}(T_p)}\right)^{1/2}
 \left({11\o n}\right)^{1/2}
 \nonumber\\[-2pt]
 &\hspace{1.0cm}\times
 \left({m_R\o230\EV}\right)^{1/2}
 \left({T\o100\GEV}\right)^{1/2}.
 \laq{Mboundnumerical}
\end{align}
Thermal equilibration then requires $y_\nu\gtrsim10^{-3}$,
where $y_\nu^2\equiv\sum_\alpha|y_{\alpha\nu}|^2$.
Because the present value $m_M(0)$ is smaller than its value at high
temperature, this one-flavor sterile-neutrino extension is in tension with
neutrino-mass and big-bang-nucleosynthesis constraints~\cite{Boyarsky:2020dzc} once the thermal-force bound is imposed.

\paragraph{Phenomenologically viable $\nu$ model}
The phenomenological problem of the one-flavor model can be avoided by
introducing a second right-handed neutrino, $\nu_{R,2}$, with
an off-diagonal mass and a negligible Yukawa coupling to the
active-lepton sector.
This is a natural extension, since neutrino masses
generated by multiple right-handed neutrinos can explain neutrino
oscillations; see Appendix~\ref{app:twoRHNspectrum}.
\begin{equation}
 {\cal L}\supset-M\nu_{R,1}\nu_{R,2}+{\rm h.c.},
 \qquad 1\GEV<M<100\GEV.
 \laq{twoRHNmassterm}
\end{equation}
The two states form a pseudo-Dirac pair. 
The constant mass makes both
sterile states heavy even when the $P$-dependent Majorana mass is small.
 If the small
$\U(1)$-violating parameters described in Appendix~\ref{app:twoRHNspectrum}
are also set to zero, lepton number is restored and the active-neutrino mass
vanishes, whereas the right-handed neutrinos remain heavy.  Nonzero
active-neutrino masses are generated by these small $\U(1)$-violating
parameters.

When both pseudo-Dirac states are thermalized,
the leading high-temperature free energy relevant to $P$ is again proportional to
$T^2\operatorname{Tr}({\cal M}^\dagger{\cal M})/24
=T^2(2M^2+m_M^2)/24$.  The $P$-dependent radial force is
therefore
$\partial_R\Delta V_T=(n/12)T^2m_M^2/R$, which leads to
Eq.~\eqref{eq:Mboundnumerical}.

\begingroup
The same thermal-force bound also suppresses thermal production of 
$P$ quanta.  Expanding the $P$-dependent Majorana mass about the rotating
background gives $g_{r\nu\nu}=|\partial_Rm_M|=nm_M/R$.  Parametrizing the
thermal collision kernel by $C_P$, one obtains
\begin{equation}
 \Gamma_P^{\rm th}=C_P
 \left({n m_M(T)\over R(T)}\right)^2T
 \ll 12C_Pn{m_R^2\over T}.
 \laq{Pthermalproductionbound}
\end{equation}
At $T=T_{\rm sph}$, this gives
\begin{equation}
 \begin{aligned}
 {\Gamma_P^{\rm th}\over H}&\ll2.14\,C_P
 \left({n\over11}\right)
 \left({m_R\over230\EV}\right)^2\\[-2pt]
 &\quad\times
 \left({131.7\GEV\over T}\right)^3
 \left({110.25\over g_*(T)}\right)^{1/2},
 \end{aligned}
 \laq{Pthermalproductionnumerical}
\end{equation}
Since $C_P \ll 1/(8\pi)$, this interaction does not thermalize $P$
quanta, and the small thermally produced population does not constitute
ordinary hot dark matter.  In the free quadratic $P$ theory, the excited and
homogeneous modes do not mix.  Thus, the small population of produced
$\delta\theta$ quanta does not contribute to the effective mass of
$R$,\footnote{This can be checked explicitly from the fact that the term
$\vev{\partial_\mu\delta\theta\,\partial^\mu\delta\theta}R$ in the equation of
motion for $R$ vanishes.} preserving the coincidence relation.
The Coleman--Weinberg correction within this model is also neglected
under the thermal-force bound~\cite{Coleman:1973jx}; its stability can be
further ensured in ultraviolet completions (see Sec.~\ref{chap:discussion}).
\endgroup

{We assume that the $P$-dependent contribution to the active-neutrino
mass is subdominant or predominantly associated with the lightest neutrino, whose
mass is small, so that the $P$ dark matter
does not decay too rapidly into
active neutrinos; see Appendix~\ref{app:twoRHNspectrum}. 
}

The change relative to the one-flavor model is the condition for
chemical equilibration.  We evaluate it using the two-state quantum
kinetic equation described in Appendix~\ref{app:twoRHNtransport}.

Figure~\ref{fig:twoRHNparameterregion} shows the parameter region
obtained from the constraints discussed above for
$M=10\GEV$ and $R(T_{\rm sph})=10^4\GEV$.  We define
$ y_{R,\mathrm{eff}}(T_{\rm sph})\equiv \sqrt{n}\,|y_P|
 \left({R(T_{\rm sph})/\sqrt{2}}\right)^{n-1}.
$
The dimensionless horizontal coordinate can therefore be used for any
integer $n$, and the radial thermal-force boundary is independent of $n$ in
this normalization.
The DELPHI active--sterile-mixing limit corresponds to
$y_\nu<2.6\times10^{-4}$ through
$|U_{\alpha N}|\simeq |y_{\alpha\nu}|v/(\sqrt{2}M)$ in the convention used
here~\cite{DELPHI:1996qcc}.  Depending on the active flavor coupled to
$\nu_{R,1}$, stronger limits may apply; see, for example,
Ref.~\cite{CMS:2023jqi}.  A sizable parameter region therefore remains viable.

\begin{figure*}[t]
 \centering
 \includegraphics[width=0.88\textwidth]{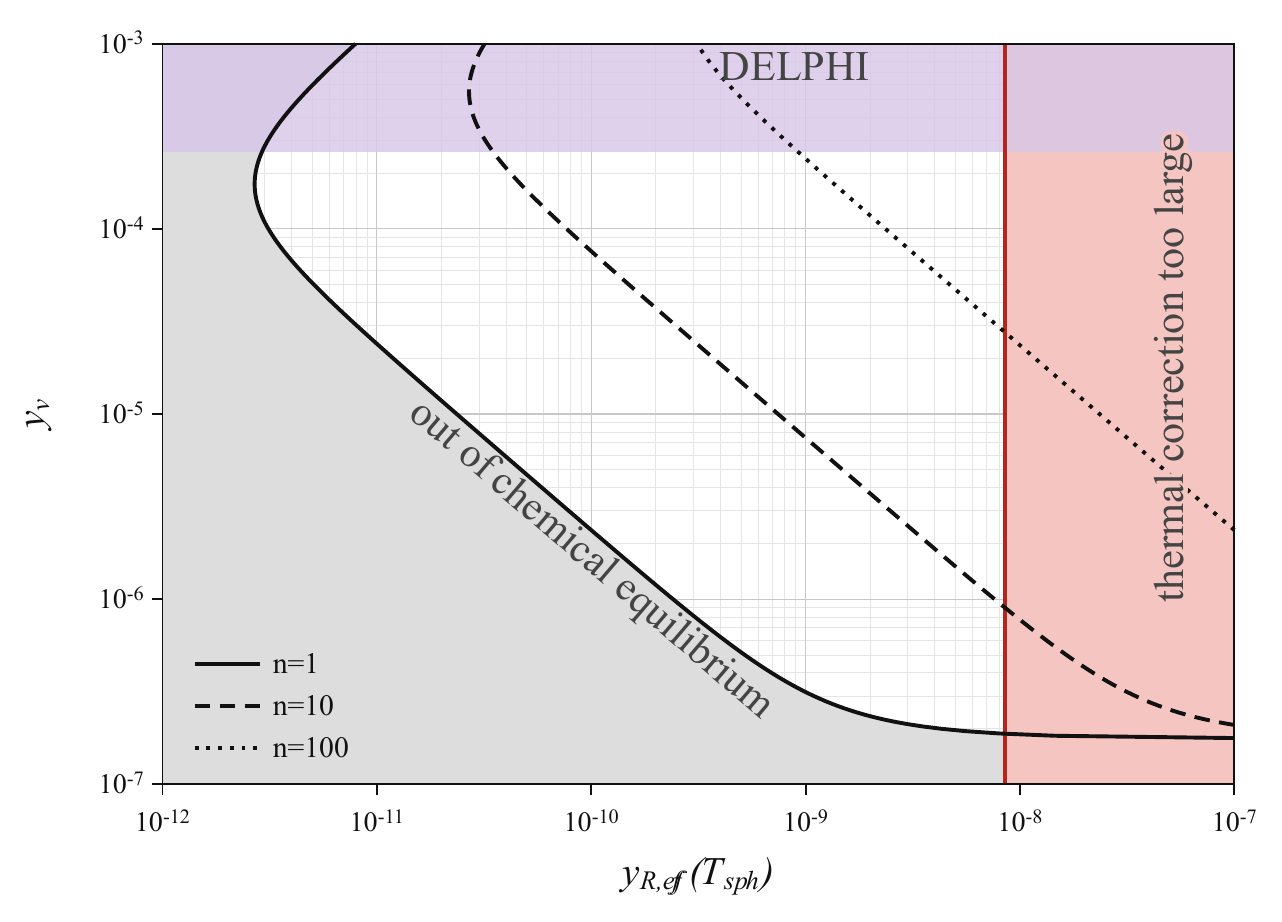}\vspace{-0.5cm}
 \caption{Parameter region in the
 $(y_{R,\mathrm{eff}}(T_{\rm sph}),y_\nu)$ plane for
 $T_{\rm sph}=131.7\GEV$, $M=10\GEV$, $R(T_{\rm sph})=10^4\GEV$, and
 $m_R=230\EV$.   The chemical-equilibration boundaries are obtained by
 equating the QKE transfer rate derived in
 Appendix~\ref{app:twoRHNtransport} to the Hubble rate.  The black solid,
 dashed, and dotted curves correspond to $n=1$, $10$, and $100$,
 respectively.
The out-of-chemical-equilibrium region and the region with excessively
large thermal corrections are also shown.
 The purple band, $y_\nu>2.6\times10^{-4}$, is excluded by the DELPHI
 active--sterile-mixing limit.}
 \label{fig:twoRHNparameterregion}
\end{figure*}

\section{Discussion}
\lac{discussion}
\subsection{Naturalness}
The model requires a CP-even scalar with a mass in the eV--keV range,
whose stability against radiative corrections calls for an organizing
principle.  One possibility is that $P$ is a complex
pseudo-Nambu--Goldstone field associated with the spontaneous symmetry
 breaking $\SU(2)\to\U(1)$.  Its small mass
and interactions can then be controlled by explicit $\SU(2)$
breaking that preserves the residual $\U(1)$.  Yukawa couplings to fermions
may preserve this $\U(1)$ while explicitly breaking the parent $\SU(2)$; see
Refs.~\cite{Sakurai:2021ipp,Haghighat:2022qyh} for a CP-even ALP with Higgs-like Yukawa
interactions.

Another possibility is a feebly interacting model with a large
wave-function-renormalization factor for
$P$~\cite{Yin:2024txg,Yin:2024pri}.  Before
canonical normalization, the Lagrangian need not contain parametrically small
coefficients; canonical normalization then makes the physical couplings weak
and the scalar mass naturally small. 

In both cases, the small mass and couplings can be protected against
radiative corrections. 
  The desired potential is recovered over the field
range in which masses induced by higher-order terms in the $P$ potential are
negligible. 
These possibilities are consistent with the parameter region because
$y_{R,\mathrm{eff}}$ is small.  For example, for $n=1$,
$y_P\sim10^{-10}$ in Fig.~\ref{fig:twoRHNparameterregion} is compatible with
$m_R=1\TEV$ before wave-function renormalization in the feebly interacting
scenario~\cite{Yin:2024txg,Yin:2024pri}, for which higher-order terms in the
potential are negligible when $R\ll10^{13}\GEV$.

\subsection{Explaining the coincidence through the attractor}
Using Eq.~\eqref{eq:RHNBaryonDensity}, whose numerical coefficient is
smaller than unity, we obtain
\begin{equation}
\begin{aligned}
\frac{\Omega_X}{\Omega_b}&\simeq
5.4\,\frac{11g_{\chi_2}}{n}
\frac{g_{*s}(T_{\rm sph})}{g_{*s}(T_p)}\\[-2pt]
&\quad\times
\left(\frac{T_{\rm sph}}{131.7\,{\rm GeV}}\right)
\left(\frac{0.94\,{\rm GeV}}{m_N}\right).
\end{aligned}
 \laq{partialburstratio}
\end{equation}
This result depends neither on the dark-matter mass nor on the
microscopic coupling, but mainly on the relevant numbers of degrees of
freedom.  This simple attractor scenario, however, leaves a factor-of-eleven
gap to be filled.

\paragraph{Full-burst and large $n$}
Consider the benchmark $\kappa_P\simeq0.09$ with one $\chi_2$
degree of freedom, obtained from burst production through stimulated emission
in $\chi_1\to\chi_2+P$.  In the full-burst initial condition, the excited
$\chi_2$ and $P$ populations are absent, although a $P$ condensate may already
be present.  For the final state generated by thermal $\chi_1$ decay, the
equilibrium result favors a modest integer charge $n=\mathcal O(10)$.

 As one explicit full-burst benchmark, taking
$g_{\chi_2}=1$, $n=11$, and
$g_{*s}(T_{\rm sph})/g_{*s}(T_p)=1$ gives
$\Omega_P/\Omega_b\approx 5.4$, in agreement with the measured value.
 This
can be realized in a UV-complete model similar to a clockwork mechanism~\cite{Choi:2015fiu,Kaplan:2015fuy,Higaki:2015jag,Giudice:2016yja} by
integrating out three additional complex scalars; see
Appendix~\ref{app:n11completion}.

\paragraph{Little burst}
More generally, burst production need not begin with an empty
$\chi_2$ sector.  If $\chi_2$ is already populated while the excited $P$
modes are absent, the number of produced $P$ quanta is suppressed relative to
the full-burst case.  In particular, if $\chi_2$ is an ordinary thermal relic,
the final $P$ abundance is the difference between two attractor abundances and
is therefore itself fixed by an attractor.
  For the
partial burst described in Appendix~\ref{app:coldhot}, write
$\kappa_P^{\rm part}\simeq0.09g_{\chi_2}^{\rm eff}$.  The equilibrium
result is obtained from Eq.~\eqref{eq:partialburstratio} by replacing
$g_{\chi_2}$ with $g_{\chi_2}^{\rm eff}$.

Consequently, $n=1$ with
$g_{\chi_2}^{\rm eff}\simeq1/11$ reproduces the same ratio, and $n=11$ is
not required.  At fixed $g_{*s}(T_p)$, the smaller partial-burst abundance
raises the dark-matter mass: this $n=1$ example gives
$m_R\simeq2.5\KEV\,[g_{*s}(T_p)/110]$.  Thus a production sector with
$g_{*s}(T_p)=\mathcal O(1)$ still gives an eV-scale mass. 

\subsection{Explaining the coincidence through deviations from the attractor}
The factor-of-ten gap can also be bridged with $n=1$ by slightly
relaxing the assumptions made above.  For instance, one may consider boundary
effects near the thermal-correction boundary shown in
Fig.~\ref{fig:twoRHNparameterregion}, where the thermal mass can enhance
baryon production during the electroweak crossover.
Taking such boundary effects into account, the baryon asymmetry can be
explained for $m_R=1\EV$--$1\KEV$.\footnote{The correct value may also be
obtained by delaying sphaleron decoupling to a lower temperature, or by
introducing a baryon-number-violating sector whose bias remains active to a
lower temperature.}

Assuming burst production, the thermal-scale dark-matter relic is also
predicted to lie in the eV--keV mass range, depending on the effective number
of degrees of freedom at production.  In particular, reheating may originate
in the right-handed-neutrino sector, which contains relatively few
relativistic degrees of freedom in our model, and be followed by thermalization
with the Standard Model.  The reheating or subsequent thermalization process
can be accompanied by burst production, in which case the suppressed
$g_{*s}(T_p)=\O(1)$ predicts $m_R\simeq1\EV$ for the full-burst
case.

{So far, we have considered only a radiation-dominated epoch.
Late-time entropy production after baryogenesis and dark-matter production
does not change the coincidence as shown in \Sec{general}, but it allows
a larger dark-matter mass, up to the MeV scale.  In this case, conventional
warm dark matter in the mass range from a representative lower bound of
$5.3\KEV$ under the fiducial smooth thermal-history assumption~\cite{Irsic:2017ixq}
to the MeV scale, without
burst production, can also explain the coincidence.}

Given the mass range, proposed experiments such as the NIRSpec infrared
spectrograph at JWST, WINERED at the Magellan Telescope, laser colliders, WISP
searches at synchrotron-radiation facilities~\cite{Bessho:2022yyu,Homma:2022ktv,Yin:2024rjb,Yin:2025bui},
and X- and gamma-ray searches may probe not only the dark-matter candidate but
also the origin of the baryon asymmetry and the reason for the coincidence
between the dark-matter and baryon abundances.
\section*{Acknowledgments}
W.Y. would like to thank the organizers of Progress in Particle
Physics 2026 and Anson Hook for discussions of his studies in
Refs.~\cite{Brzeminski:2023wza,Banerjee:2024xhn}, which prompted W.Y. to
reconsider an idea first explored in 2020 and ultimately led to this work.
W.Y. is supported by JSPS KAKENHI Grant Nos. 22K14029, 23K22486, and 26K00695 and by the Selective Research Fund and Incentive Research Fund from Tokyo Metropolitan University.
\appendix

\section{Burst production of cold ``hot dark matter'' via stimulated emission}
\label{app:coldhot}
\subsection{Review of burst production}
Once quantum statistics is included, the thermal production of a light
boson can behave analogously to a laser: Bose-enhanced stimulated emission
rapidly amplifies infrared modes until the system reaches a quasi-steady
state~\cite{Yin:2023jjj,Sakurai:2024apm}.  As an explicit example, consider
\begin{equation}
 \chi_1\leftrightarrow\chi_2+X,
 \qquad T_p\gg M_1>M_2\gg m_X.
 \laq{decayprocess}
\end{equation}
Here $\chi_{1,2}$ are generic particles.  We take $\chi_1$ to have a
nearly thermal distribution, while $\chi_2$ and $X$ are initially absent.
Defining
\begin{equation}
 \y=1-{M_2^2\o M_1^2},
 \laq{eta}
\end{equation}
a backward-emitted $X$ from the decay of a boosted $\chi_1$ has the
characteristic momentum
\begin{equation}
 p_X^{\rm burst}\simeq\y{M_1^2\o2T_p}\ll T_p.
 \laq{pburst}
\end{equation}
This soft mode is populated rapidly because its phase-space volume is
proportional to $(p_X^{\rm burst})^3$, and the corresponding occupation number
is inversely proportional to that volume.  The inverse timescale for the
occupation number to reach unity is
\begin{equation}
 \G_{\rm ign}\sim {g_{\chi_1}\o g_X}
 {4T_p^3\o\y^3M_1^3}\G^{\rm rest}_{\chi_1\to\chi_2X},
 \laq{ignition}
\end{equation}
Here, the suppression from the small phase space is compensated by the
correspondingly small phase-space volume entering the occupation number.
The usual thermally averaged decay rate 
\begin{equation}
 \G_{\rm dec}^{\rm th}\sim
 \G^{\rm rest}_{\chi_1\to\chi_2X}{M_1\o T_p},
 \laq{thermaldecay}
\end{equation}
is smaller. 

Once the Bose-enhanced burst begins, the infrared modes grow
exponentially.  Backreaction through the inverse decay terminates this growth
when $f_{\chi_2}(p_{\chi_2}\sim T_p)=\mathcal{O}(1)$, since
$p_{\chi_2}+p_{X}^{\rm burst}\sim p_{\chi_1}\sim T_p$.
Using
$n_{\chi_2}=n_X$, one obtains
\begin{equation}
 \begin{aligned}
 n_X^{\rm burst}(p\sim p_X^{\rm burst})
 &=n_{\chi_2}(p\sim T_p),\\
 n_{\chi_2}(p\sim T_p)&\equiv\k_X^{\rm burst}T_p^3,\\
 \k_X^{\rm burst}&\simeq {g_{\chi_2}\o\pi^2}.
 \end{aligned}
 \laq{burstnumber}
\end{equation}
Numerical simulations give
$\k_X^{\rm burst}\approx0.09g_{\chi_2}$~\cite{Yin:2023jjj}.
For a single effective degree of freedom, a natural benchmark is therefore
$\k_X\sim0.09$.  {Within the regime specified below, this
quasi-equilibrium state is an attractor, and its abundance is insensitive to
variations in $\G_{\rm ign}$ and $\G_{\rm dec}^{\rm th}$.}  The regime is
\begin{equation}
 \G_{\rm ign}\gg H\gg\G_{\rm dec}^{\rm th}.
 \laq{burstcondition}
\end{equation}
Here $H$ is the Hubble expansion rate.  This hierarchy permits a rapid
infrared burst without ordinary thermalization.  The same
$1\leftrightarrow2$ interaction subsequently becomes kinematically
ineffective as the Universe expands, and the cold $X$ population
free-streams.  Such a hierarchy is readily realized for weakly coupled bosons and
other WISPs~\cite{Jaeckel:2010ni,Ringwald:2012hr,Arias:2012az,Graham:2015ouw,Marsh:2015xka,Irastorza:2018dyq,DiLuzio:2020wdo,Albertus:2026fbe,Arza:2026rsl}.  Bose-enhanced production can also occur in
$2\to2$ scattering dominated by soft momentum transfer~\cite{Sakurai:2024apm}.

\subsection{A little burst}
The abundance need not correspond to a full burst into an initially
empty $\chi_2$ sector; the partial-burst solution can still be an attractor.
 Suppose instead that $\chi_1$ and $\chi_2$ were
initially in equilibrium, after which $\chi_2$ decoupled and entropy release
heated the $\chi_1$ bath, as in standard cosmology after neutrino decoupling.
 At production, let $T_{\chi_1}=T_p$ and
$T_{\chi_2}=T'=\xi T_p$, where entropy conservation in the $\chi_1$ sector
gives
\begin{equation}
 \xi\equiv{T'\over T_p}
 =\left({g_{*s}^{\rm after}\over g_{*s}^{\rm before}}\right)^{1/3}<1.
 \laq{partialbursttemperatureratio}
\end{equation}
If $X$ is initially absent, the stimulated process fills only the
deficit in the $\chi_2$ distribution.  Parametrically,
\begin{equation}
 \begin{aligned}
 n_X^{\rm part}=\Delta n_{\chi_2}
 &\simeq\kappa_X^{\rm burst}T_p^3(1-\xi^3),\\
 \kappa_X^{\rm part}&\simeq0.09g_{\chi_2}^{\rm eff},
 \end{aligned}
 \laq{partialburstnumber}
\end{equation}
\begin{equation}
 g_{\chi_2}^{\rm eff}\equiv g_{\chi_2}(1-\xi^3)
 =g_{\chi_2}{g_{*s}^{\rm before}-g_{*s}^{\rm after}
 \over g_{*s}^{\rm before}}.
 \laq{partialbursteffectiveg}
\end{equation}
Thus the integer multiplicity $g_{\chi_2}$ is unchanged, whereas the
effective abundance factor is continuous and can be
$\mathcal O(10^{-2}\text{--}10^{-1})$ for $g_{\chi_2}=\O(1)$ when the entropy degrees of
freedom decrease by $1$--$10\%$.  The numerical
coefficient above is a parametric rescaling of the full-burst result; a
dedicated kinetic calculation with this initial condition is left for future
work.

\section{Chemical bias and thermal transport}
\label{app:twoRHNtransport}

We write the rotating Majorana mass as
\begin{equation}
 m_M(P)=m_M(R)e^{in\theta},
 \qquad \omega=n\dot\theta.
\end{equation}
The field redefinition
\begin{equation}
 \nu_{R,1}=e^{-in\theta/2}\widetilde\nu_{R,1}
\end{equation}
removes the phase from the diagonal Majorana entry and generates
\begin{equation}
 {\cal L}\supset{n\over2}\partial_\mu\theta\,
 \overline{\widetilde\nu_{R,1}}\gamma^\mu\widetilde\nu_{R,1}.
 \laq{twoRHNderivativesource}
\end{equation}
This is the Majoron-type derivative-coupling bias used in spontaneous
leptogenesis~\cite{Li:2001st,Kusenko:2014lra,Kusenko:2014uta,Ibe:2015nfa}.
Compensating phases then appear in the Yukawa and off-diagonal mass
terms, so the equilibrium relations are most simply imposed in the original
basis.  When the Majorana and Yukawa reactions in
Eq.~\eqref{eq:lagrangian} and the off-diagonal mass interaction in
Eq.~\eqref{eq:twoRHNmassterm} are in equilibrium,
\begin{equation}
 2\mu_{\nu_{R,1}}=\omega,
~
 \mu_{\nu_{R,1}}=\mu_\ell+\mu_H,
 ~
 \mu_{\nu_{R,2}}=-\mu_{\nu_{R,1}}.
 \laq{twoRHNsourceequilibrium}
\end{equation}
The Standard Model Yukawa-equilibrium conditions, the weak-sphaleron
condition, and the hypercharge-neutrality condition are
\begin{align}
 \mu_u=\mu_q+\mu_H,~
 \mu_d=\mu_q-\mu_H,~
 \mu_e=\mu_\ell-\mu_H,\non \\
 3\mu_q+\mu_\ell=0,~
 6\mu_q-6\mu_\ell+14\mu_H=0.
 \laq{twoRHNSMequilibrium}
\end{align}
Because the Majorana operator violates $B-L$, no additional $B-L$
conservation equation is imposed.  Equations
\eqref{eq:twoRHNsourceequilibrium} and
\eqref{eq:twoRHNSMequilibrium} give
\begin{equation}
 \mu_q=-{7\over66}\omega,
 \qquad \mu_\ell={7\over22}\omega,
 \qquad \mu_H={2\over11}\omega.
\end{equation}
Since $n_B=2\mu_qT^2$ for three relativistic Standard Model generations,
this gives Eq.~\eqref{eq:RHNBaryonDensity}.  
\begingroup

For a controlled treatment away from instantaneous equilibrium, we introduce a
helicity-resolved $2\times2$ density matrix $\rho_h(k)$.  To leading order in
the sterile-neutrino Yukawa coupling, its quantum kinetic equation (QKE) has the standard
commutator--anticommutator form
\begin{equation}
 \dot\rho_h=-i[H_h,\rho_h]
 -{1\over2}\{\Gamma_h,\rho_h-\rho_h^{\rm eq}\}+{\cal S}_h,
 \laq{twoRHNdensitymatrix}
\end{equation}
where ${\cal S}_h$ contains the bias and plasma-charge backreaction.  The
source fixes the stationary solution but does not change the eigenvalues of
the homogeneous linearized equation.  Equations of this form, including
helicity and momentum dependence, are derived in
Refs.~\cite{Sigl:1993ctk,Drewes:2016gmt,Hamada:2018epb,Ghiglieri:2017gjz,Yin:2024trc}.

For the number-density equation, we adopt the usual momentum average
\begin{equation}
 \left\langle{1\over k}\right\rangle
 \equiv{\displaystyle\int{d^3k\over(2\pi)^3}
 {f_F(k)\over k}\over
 \displaystyle\int{d^3k\over(2\pi)^3}f_F(k)}
 ={\pi^2\over18\zeta(3)T}\equiv{1\over\bar E},
 \laq{twoRHNmomentumaverageAppendix}
\end{equation}
as in Ref.~\cite{Drewes:2016gmt}.  Define
$\omega\equiv n\dot\theta$ and
$y_\nu^2\equiv\sum_\alpha|y_{\alpha\nu}|^2$.  The damping rate is
\begin{equation}
 \Gamma_Y(T)=\gamma_{\rm av}y_\nu^2T,
 \qquad D={\Gamma_Y\over2},
 \qquad \gamma_{\rm av}\simeq0.012,
 \laq{twoRHNdamping}
\end{equation}
in the one-sided damping approximation
\cite{Drewes:2016gmt,Ghiglieri:2017gjz}.  The oscillation frequency and the
helicity-dependent detuning are
\begin{equation}
 \begin{aligned}
 \Delta E&={Mm_M(T)\over\bar E},
~~~~~~~\Delta V_Y={y_\nu^2T^2\over8\bar E},\\
 \delta_h&=\Delta V_Y-h\omega+{m_M(T)^2\over2\bar E},
 &h&=\pm1.
 \end{aligned}
 \laq{twoRHNdetuning}
\end{equation}
In the corotating interaction basis, after dropping a term proportional
to the identity, the Hamiltonian and
damping matrix are
\begin{equation}
 H_h={1\over2}
 \begin{pmatrix}
  \delta_h&\Delta E\\
  \Delta E&-\delta_h
 \end{pmatrix},
 \qquad
 \Gamma_h=\begin{pmatrix}\Gamma_Y&0\\0&0\end{pmatrix},
 \laq{twoRHNHamiltonian}
\end{equation}
Here $h=\pm1$ labels the helicity.
The term
$y_\nu^2T^2/(8\bar E)$ in $\delta_h$ is a dispersive thermal potential and
is distinct from the $P$-dependent thermal free energy in
Eq.~\eqref{eq:thermalmassbound}
\cite{Drewes:2016gmt,Ghiglieri:2016xye,Ghiglieri:2017gjz}.  The rotation
frequency $\omega=n\dot\theta$ therefore enters the transfer rate through
the detuning, even though the amplitude of the constant source does not
multiply a linear relaxation eigenvalue.

Let $\rho_h^{\rm ss}$ denote the stationary solution of
Eq.~\eqref{eq:twoRHNdensitymatrix} and write
\begin{equation}
 \delta\rho_h\equiv\rho_h-\rho_h^{\rm ss}
 =\begin{pmatrix}f_1&a+ib\\a-ib&f_2\end{pmatrix}.
\end{equation}
For fixed $T$ and $R$, the homogeneous QKE becomes
\begin{equation}
 {d\over dt}
 \begin{pmatrix}f_1\\f_2\\a\\b\end{pmatrix}
 =
 \begin{pmatrix}
 -2D&0&0&-\Delta E\\
 0&0&0&\Delta E\\
 0&0&-D&\delta_h\\
 \Delta E/2&-\Delta E/2&-\delta_h&-D
 \end{pmatrix}
 \begin{pmatrix}f_1\\f_2\\a\\b\end{pmatrix}.
 \laq{twoRHNlinearQKE}
\end{equation}
This equation also shows explicitly the coherent transfer through the
off-diagonal density-matrix element.  For example, the coherence satisfies
\begin{equation}
 \dt{\delta\rho}_{12}=-(D+i\delta_h)\delta\rho_{12}
 -i{\Delta E\over2}(f_2-f_1).
 \laq{twoRHNcoherence}
\end{equation}

The $4\times4$ system can be solved analytically.  Write a decay eigenvalue
as $-\Gamma$, and define
\begin{equation}
 u\equiv1-{\Gamma\over D},
 \qquad X\equiv{\Delta E\over D},
 \qquad Z_h\equiv{\delta_h\over D}.
\end{equation}
The characteristic equation reduces to
\begin{equation}
 u^4+(X^2+Z_h^2-1)u^2-Z_h^2=0.
 \laq{twoRHNcharacteristic}
\end{equation}
The root closest to zero decay gives
\begin{align}
 U_h^2&={1-X^2-Z_h^2+
 \sqrt{(X^2+Z_h^2-1)^2+4Z_h^2}\over2},
 \nonumber\\
 \Gamma_h^{\rm QKE}&=D
 \begin{cases}
  1-\sqrt{U_h^2},&U_h^2\geq0,\\
  1,&U_h^2<0,
 \end{cases}
 \laq{twoRHNgapsolution}
\end{align}
The two helicity blocks are independent at this order, so the spectral gap of
the combined sterile system is $\min_h\Gamma_h^{\rm QKE}$.  Expanding
Eq.~\eqref{eq:twoRHNgapsolution} for
$X^2\ll1+Z_h^2$ gives
\begin{equation}
 \Gamma_h^{\rm QKE}\simeq {\Delta E^2D\over
 2(D^2+\delta_h^2)}.
 \laq{twoRHNweakgapAppendix}
\end{equation}
  At exact resonance, $Z_h=0$, the gap becomes
$D[1-\sqrt{1-X^2}]$ for $X<1$ and saturates at $D$ for $X\geq1$.

As a direct numerical check of the overdamped approximation,
Fig.~\ref{fig:twoRHNtimeevolution} shows the solution of
Eq.~\eqref{eq:twoRHNlinearQKE} for an initial perturbation
$\delta\rho_h(0)={\rm diag}(0,1)$ at an overdamped point satisfying
$X^2\ll1+Z_h^2$.  The solution rapidly projects onto the slow eigenmode, and
its decay is reproduced by Eq.~\eqref{eq:twoRHNweakgapAppendix}.

\begin{figure*}[t]
 \centering
 \includegraphics[width=0.58\textwidth]{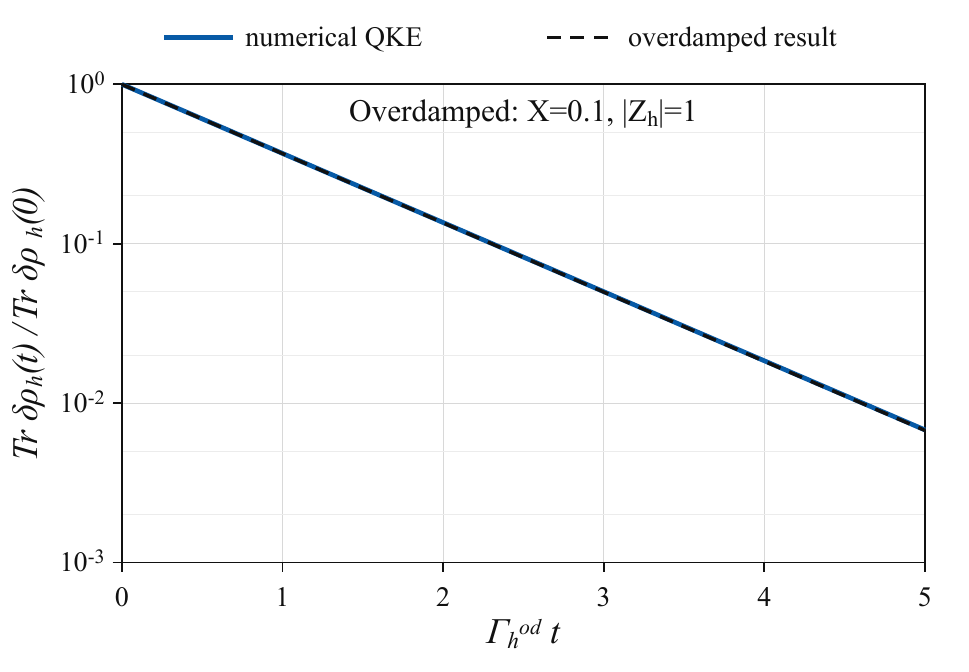}
 \caption[Numerical evolution of the homogeneous QKE in the overdamped
 regime.]{{Numerical evolution of the homogeneous QKE in
 Eq.~\eqref{eq:twoRHNlinearQKE} at the overdamped point
 $X=0.1$ and $|Z_h|=1$, starting from
 $\delta\rho_h(0)={\rm diag}(0,1)$.  The blue curve is the full
 matrix-exponential solution, and the black dashed curve is the overdamped
 result in Eq.~\eqref{eq:twoRHNweakgapAppendix}.}}
 \label{fig:twoRHNtimeevolution}
\end{figure*}

Equation~\eqref{eq:twoRHNgapsolution} is exact for the momentum-averaged
relaxation-time QKE in Eq.~\eqref{eq:twoRHNdensitymatrix}.  Momentum-dependent
collision kernels, the second small Yukawa coupling, and the electroweak
crossover generate order-one corrections, which can be incorporated by
solving the full QKE and plasma-charge system
\cite{Eijima:2017anv,Ghiglieri:2017gjz,Laine:2022pgk,Hernandez:2022ivz}.

\section{Two right-handed neutrinos and the light spectrum}
\label{app:twoRHNspectrum}

The Yukawa interactions of the two right-handed neutrinos are
\begin{equation}
 \begin{aligned}
 \mathcal{L}_Y\supset{}&-\sum_\alpha y_{\alpha\nu}
 \overline{L_\alpha}\widetilde H\nu_{R,1}\\
 &-\sum_\alpha y_{\alpha2}
 \overline{L_\alpha}\widetilde H\nu_{R,2}+{\rm h.c.}
 \end{aligned}
 \laq{twoRHNYukawas}
\end{equation}
After electroweak symmetry breaking,
$m_{D1,\alpha}=y_{\alpha\nu}v/\sqrt{2}$ and
$m_{D2,\alpha}=y_{\alpha2}v/\sqrt{2}$, where $v=246\GEV$.  We define
$y_\nu^2\equiv\sum_\alpha|y_{\alpha\nu}|^2$ and
$y_2^2\equiv\sum_\alpha|y_{\alpha2}|^2$.
Because $y_2\ll y_\nu$ in the benchmark below, its diagonal damping
and dispersive thermal potential are neglected in the main-text QKE.  We
also assume that the flavor overlap
$\sum_\alpha y_{\alpha\nu}^*y_{\alpha2}$ is small enough for the
off-diagonal thermal terms to be negligible for simplicity.
We denote the resulting Dirac-mass flavor vectors by $m_{D1}$ and $m_{D2}$.
We also allow a small, $P$-independent Majorana mass for
$\nu_{R,2}$,
\begin{equation}
 {\cal L}\supset-{1\over2}\mu_2
 \overline{\nu_{R,2}^{\,c}}\nu_{R,2}+{\rm h.c.},
 \qquad |\mu_2|\ll M.
 \laq{twoRHNmu2}
\end{equation}
The $y_2$ coupling (equivalently, $m_{D2}$) and a nonzero $\mu_2$
explicitly break the $\U(1)$ symmetry, whereas $m_{D1}$ preserves it.
For the cosmological analysis above, we assume
$|\mu_2|\ll|m_M(T_{\rm sph})|,\qquad y_2\ll y_\nu$, so that
their effects do not modify the QKE or the chemical-equilibrium conditions.
To leading order in $m_{Di}/M$, $m_M/M$, and $\mu_2/M$, integrating out $\nu_{R,1}$ and $\nu_{R,2}$ gives
\begin{equation}
 \boxed{\begin{aligned}
 m_\nu={}&-{m_{D1}m_{D2}^T+m_{D2}m_{D1}^T\over M}\\
 &+{\mu_2\over M^2}m_{D1}m_{D1}^T
 +{m_M(0)\over M^2}m_{D2}m_{D2}^T.
 \end{aligned}}
 \laq{twoRHNlightmass}
\end{equation}
For generality, we retain $m_M(0)$, which is induced by the late-time
$P$ background and vanishes for $R_0=0$ (or $y_P=0$); it is not an
explicit $\U(1)$-breaking spurion.
Two nonparallel flavor vectors generically give two nonzero light-neutrino
masses, while the third neutrino is massless.  This is the minimal
two-right-handed-neutrino spectrum; related minimal and pseudo-Dirac
realizations have been studied, for example, in
Refs.~\cite{Frampton:2002qc,Rink:2016knw}.

The PMNS matrix can be realized explicitly using a simple special
case of the master parametrization for Majorana-neutrino mass
models~\cite{Cordero-Carrion:2019qtu}; see also the Casas--Ibarra
parametrization~\cite{Casas:2001sr}.  Let
$\bm y_\nu=(y_{e\nu},y_{\mu\nu},y_{\tau\nu})^T$ and
$\bm y_2=(y_{e2},y_{\mu2},y_{\tau2})^T$, and let $\bm u_i$ denote column
$i$ of $U_{\rm PMNS}^*$.  In the normal ordering,
define $S_{\rm NO}\equiv m_2+m_3$.  When the linear-seesaw term in
Eq.~\eqref{eq:twoRHNlightmass} dominates, the choice
\begin{equation}
 \begin{aligned}
 \bm y_\nu&={y_\nu\over\sqrt{S_{\rm NO}}}
 \left(\sqrt{m_2}\,\bm u_2+i\sqrt{m_3}\,\bm u_3\right),\\
 \bm y_2&=-{M\sqrt{S_{\rm NO}}\over y_\nu v^2}\\[-2pt]
 &\quad\times
 \left(\sqrt{m_2}\,\bm u_2-i\sqrt{m_3}\,\bm u_3\right)
 \end{aligned}
 \laq{twoRHNexplicitPMNS}
\end{equation}
gives
$m_\nu=U_{\rm PMNS}^*\operatorname{diag}(0,m_2,m_3)U_{\rm PMNS}^\dagger$
exactly and fixes the norm of the second Yukawa vector to
\begin{equation}
 \begin{aligned}
 y_2&={M(m_2+m_3)\over y_\nu v^2}\\
 &\simeq9.72\times10^{-10}
 \left({M\over10\GEV}\right)
 \left({10^{-5}\over y_\nu}\right)
 \left({m_2+m_3\over0.0588\EV}\right).
 \end{aligned}
 \laq{twoRHNYukawahierarchy}
\end{equation}
For inverted ordering, the same construction uses
$(m_1,\bm u_1)$ and $(m_2,\bm u_2)$ and gives
$y_2=1.63\times10^{-9}$ at the same benchmark point.  A numerical Takagi
diagonalization of the full $5\times5$ mass matrix for
$M=10\GEV$, $y_\nu=10^{-5}$, and $\mu_2=m_M(0)=0$ gives, for normal ordering,
\begin{equation}
 (m_1,m_2,m_3)
 =(0,8.6816,50.1099)\,{\rm meV},
 \laq{twoRHNnumericalmasses}
\end{equation}
together with
$(\sin^2\theta_{12},\sin^2\theta_{13},\sin^2\theta_{23},
\delta_{\rm CP})=(0.3088,0.02248,0.470,212^\circ)$, reproducing the
\href{https://www.nu-fit.org/?q=node/309}{NuFIT 6.1 online update} of
Ref.~\cite{Esteban:2024eli}.  The active--heavy
mixing norm is $1.7\times10^{-4}$, and the resulting nonunitarity is
$\mathcal O(10^{-8})$.  Thus both mass orderings and the full PMNS matrix are
realizable, with one exactly massless light neutrino.  Their nonparallel
components are essential; exact alignment reduces the rank to one.

\begin{figure}[t]
 \centering
 \includegraphics[width=\columnwidth]{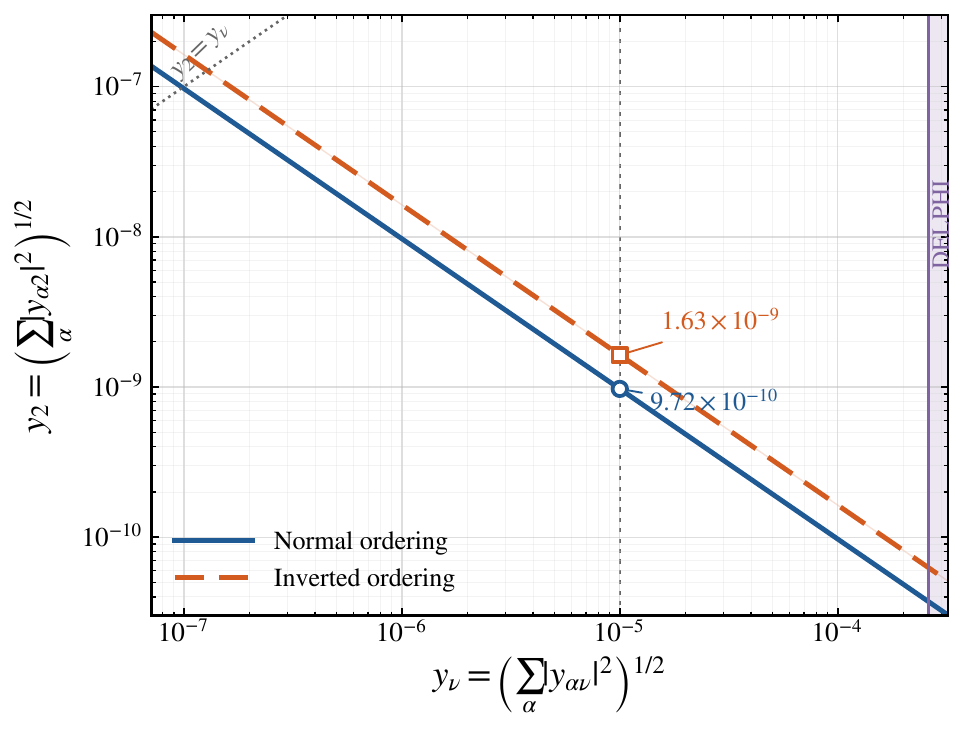}
 \caption{The Yukawa norm $y_2$ required to reproduce the two
 nonzero light-neutrino masses as a function of $y_\nu$, for $M=10\GEV$.
 The solid and dashed curves show normal and inverted ordering,
 respectively; the narrow bands use the
 \href{https://www.nu-fit.org/?q=node/309}{NuFIT 6.1 online update}
 $3\sigma$ ranges of the
 mass splittings~\cite{Esteban:2024eli}.  The markers denote
 $y_\nu=10^{-5}$, for which $y_2=9.72\times10^{-10}$ (normal ordering) or
 $1.63\times10^{-9}$ (inverted ordering).  With the flavor directions in
 Eq.~\eqref{eq:twoRHNexplicitPMNS}, every point on the curves reproduces the
 PMNS matrix.  The purple line indicates the illustrative DELPHI
 limit on $y_\nu$ used in the main text.}
 \label{fig:twoRHNpmnsfit}
\end{figure}

We next examine the decay of the complex $P$ particle into light
neutrinos.  The $P$ field is not in a spontaneously broken phase, and we
expand about $\langle P\rangle=0$.  The asymptotic scalar states are
therefore $P$ and $P^\dagger$, rather than separate radial and phase modes.  We set
$\mu_2=0$ and take $M$ to be real in the following calculation.  To
leading order in $m_{Di}/M$ and $m_M/M$, the soft, $P$-dependent
light-neutrino vertex can be written as
\begin{equation}
 \begin{aligned}
 {\cal L}_{P\nu\nu}^{\rm eff}={}&
 -{1\over2}\nu_L^TC^{-1}\\[-2pt]
 &\quad\times\big[m_M(P){\cal G}_L
 +m_M(P)^*{\cal G}_T\big]\nu_L+{\rm h.c.},\\
 {\cal G}_T&=\left({1\over M^2}+C_2(M)\right)m_{D2}m_{D2}^T,\\
 {\cal G}_L&=C_1(M)m_{D1}m_{D1}^T.
 \end{aligned}
 \laq{twoRHNeffectivePvertex}
\end{equation}
The $M^{-2}$ term in ${\cal G}_T$ is the tree-level contribution.
For $y_2=0$, the exact massless light eigenstates have no $\nu_{R,1}$
component and the tree-level vertex vanishes.  This is an alignment
property of the neutral-fermion mass matrix, not a symmetry that forbids
$P$ decay.

Finite electroweak matching generates ${\cal G}_L$ and the small
correction $C_2$ in ${\cal G}_T$.  We obtain these soft vertices by
expanding the general two-heavy-neutrino one-loop kernel of
Refs.~\cite{Dev:2012sg,Lopez-Pavon:2015cga}, rather than by
introducing an additional neutrino-mass assumption.  The relevant loop
function is
\beq
F(x)\equiv{x\over2}\left[
 3{\ln(x/m_Z^2)\over x/m_Z^2-1}
 +{\ln(x/m_h^2)\over x/m_h^2-1}\right],
 \laq{twoRHNloopfunction}
 \eeq
 \begin{align}
 C_1(M)&\equiv{2F'(M^2)\over(4\pi v)^2},
 \nonumber\\[-2pt]
 C_2(M)&\equiv{2\over(4\pi v)^2}
 \left[F'(M^2)-{F(M^2)\over M^2}\right].
 \laq{twoRHNloopcoefficients}
\end{align}
Here $m_Z$ and $m_h$ are the $Z$- and Higgs-boson masses.  The
complex conjugation of $m_M(P)$ in the ${\cal G}_T$ term of
Eq.~\eqref{eq:twoRHNeffectivePvertex} is essential for a general complex
$P$ field.  Corrections due to the external momentum are of order
$m_R^2/M^2$ and $m_R^2/m_Z^2$ and are negligible in the mass range of
interest.  For $M=10\GEV$,
\begin{equation}
 \begin{aligned}
 C_1&=1.53453\times10^{-6}\GEV^{-2},\\
 C_2&=-4.02550\times10^{-7}\GEV^{-2}.
 \end{aligned}
 \laq{twoRHNloopcoefficientsnumerical}
\end{equation}
Using $m_{D1}=v\bm y_\nu/\sqrt{2}$ and
$m_{D2}=v\bm y_2/\sqrt{2}$, it is convenient to define
$A_1\equiv v^2C_1/2=0.0464318$ and
$A_2\equiv v^2(1/M^2+C_2)/2=302.5678$.

For $n=1$, $m_M(P)=y_PP$, and
Eq.~\eqref{eq:twoRHNeffectivePvertex} becomes
\begin{equation}
 {\cal L}_{P\nu\nu}^{\rm eff}
 =-{1\over2}\nu_L^TC^{-1}
 \left(y_PP{\cal G}_L+y_P^*P^\dagger{\cal G}_T\right)\nu_L
 +{\rm h.c.}
 \laq{twoRHNcomplexPvertex}
\end{equation}
The ${\cal G}_T$ and ${\cal G}_L$ terms produce opposite-helicity
neutrino pairs and therefore do not interfere in the limit
$m_R\gg m_\nu$.  The widths of $P$ and $P^\dagger$ are equal by CPT and
are
\begin{equation}
 \begin{aligned}
 \Gamma_{P\to\nu\nu}&=\Gamma_{P^\dagger\to\nu\nu},\\
 \Gamma_{P\to\nu\nu}
 &={m_R|y_P|^2\over32\pi}\big[
 \operatorname{Tr}({\cal G}_T^\dagger{\cal G}_T)
 +\operatorname{Tr}({\cal G}_L^\dagger{\cal G}_L)\big]\\
 &={m_R|y_P|^2\over32\pi}
 \left(A_2^2y_2^4+A_1^2y_\nu^4\right).
 \end{aligned}
 \laq{twoRHNdecaywidth}
\end{equation}
The $32\pi$ denominator is the one-channel result for a complex
scalar; the two terms in Eq.~\eqref{eq:twoRHNdecaywidth} already sum the
two conjugate helicity channels available to a $P$ particle.  Neutrino-mass
corrections, including interference between the two helicity channels,
are negligible.  At the $n=1$ thermal-force boundary
$|y_P|=y_{R,{\rm eff}}=8.5\times10^{-9}$ and for $m_R=230\EV$, the
separate pieces scale as
\begin{equation}
 \begin{aligned}
 \Gamma_{\rm tree}&=2.30\times10^4 y_2^4\,{\rm s}^{-1},\\
 \Gamma_{\rm EW}&=5.41\times10^{-4} y_\nu^4\,{\rm s}^{-1}.
 \end{aligned}
 \laq{twoRHNdecayratesnumerical}
\end{equation}
For $y_\nu=10^{-5}$, the explicit PMNS textures above give
$\Gamma_{P\to\nu\nu}\simeq5.41\times10^{-24}\,{\rm s}^{-1}$ for either
mass ordering, corresponding to a lifetime of about
$1.85\times10^{23}\,{\rm s}$.

\begin{figure*}[t]
 \centering
 \includegraphics[width=0.82\textwidth]{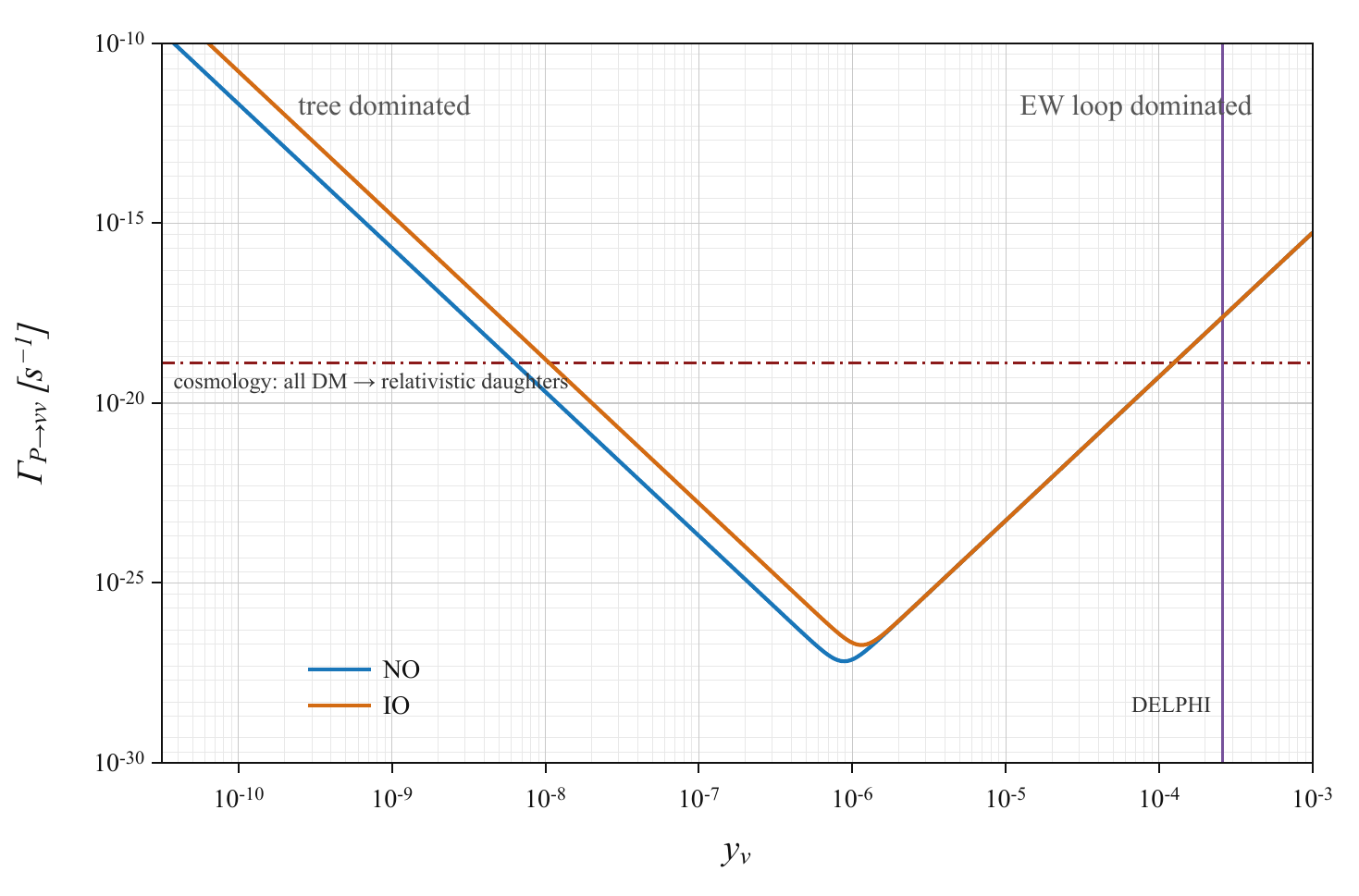}
 \caption{Thermal-force upper envelopes of the complex-$P$ decay
 rate into light neutrinos as a function of $y_\nu$, for $n=1$,
 $M=10\GEV$, and $m_R=230\EV$.  The explicit PMNS texture fixes
 $y_2=9.72\times10^{-15}/y_\nu$ for normal ordering (NO) and
 $y_2=1.63\times10^{-14}/y_\nu$ for inverted ordering (IO).  Both the
 tree and finite electroweak one-loop vertices are included.  The red
 dash-dotted horizontal line is the 95\% C.L. bound for all cold dark matter
 decaying into relativistic invisible daughters~\cite{Alvi:2022aam}.  The
 purple line is the illustrative DELPHI
 boundary used in the main text.  A smaller $|y_P|$ lowers every rate as
 $|y_P|^2$.}
 \label{fig:twoRHNdecayrate}
\end{figure*}

The cosmological analysis of Ref.~\cite{Alvi:2022aam} gives
$\Gamma_{\rm DCDM}<1.29\times10^{-19}\,{\rm s}^{-1}$, or
$\tau_{\rm DCDM}>246\,{\rm Gyr}$, at 95\% C.L. when all cold dark matter
decays into effectively massless products.  At the thermal-force envelope,
the loop-dominated branch therefore requires approximately
$y_\nu\lesssim1.24\times10^{-4}$.  The benchmark $y_\nu=10^{-5}$ is safely
long lived, but the decay is not automatically negligible over the entire
$y_\nu$ range.  The small-$y_\nu$, $y_2$-dominated region is not covered by the
main-text one-Yukawa QKE and requires a separate two-Yukawa transport
analysis.  For $n>1$ in the symmetry-preserving vacuum, the leading
operators contain $P^n$ or $(P^\dagger)^n$ and do not mediate a
single-particle $P$ decay.  A one-particle decay for $n>1$ would require
insertions of a nonzero coherent background, which is not assumed here.

A PTOLEMY-like tritium experiment can search for neutrinos from
$P\to\nu\nu$ through threshold-free neutrino capture~\cite{McKeen:2018xyz}.
The $m_R=230\EV$ case is beyond a standard exposure, whereas the capture
yield scales approximately as $N_{\rm cap}\propto1/m_R$ at fixed dark-matter
fraction and decay rate.  Consequently, for a decay rate near the present
cosmological limit, an eV-scale $P$ can be within reach, provided that the
decay is kinematically open.  

\clearpage
\section{A renormalizable integer-charge completion for $n=11$}
\label{app:n11completion}

This section gives an illustrative completion of the full-burst
$n=11$ benchmark.  The partial-burst realization does not require this
charge assignment.
The power eleven need not arise from a uniform clockwork with
$n=3^\ell$~\cite{Kaplan:2015fuy,Higaki:2015jag}.  Normalize $Q(P)=-1$ and introduce heavy complex scalars
$\Phi_{-11}$, $\phi_3$, and $\phi_4$ with the charges indicated by their
subscripts.  Taking $Q(\nu_{R,1})=11/2$ and
$Q(\nu_{R,2})=-11/2$ makes the endpoint and the constant off-diagonal mass
invariant.  Multiplying all charges by two gives an entirely integer
normalization.  The renormalizable interactions may be chosen as
\begin{align}
 {\cal L}_{\rm UV}&\supset
 -{1\over2}y_\Phi\Phi_{-11}
 \overline{\nu_{R,1}^{\,c}}\nu_{R,1}+{\rm h.c.},
 \nonumber\\
 V_{\rm UV}&\supset M_{11}^2|\Phi_{-11}|^2
 +M_3^2|\phi_3|^2+M_4^2|\phi_4|^2
 \nonumber\\
 &\quad+\left(\lambda_A\Phi_{-11}\phi_3\phi_4^2
 +\lambda_3\phi_3P^3\right.
 \nonumber\\
 &\qquad\left.+\mu_4\phi_4\phi_3^\dagger P+{\rm h.c.}\right).
 \laq{n11UVinteractions}
\end{align}
The integer charge chain is
\begin{equation}
 3=1+1+1,
 \qquad4=3+1,
 \qquad11=3+4+4.
\end{equation}
Tree-level integration of $\phi_3$, $\phi_4$, and $\Phi_{-11}$ generates
\begin{equation}
 -{1\over2}y_PP^{11}
 \overline{\nu_{R,1}^{\,c}}\nu_{R,1}+{\rm h.c.},
\end{equation}
with the parametric matching
\begin{equation}
 |y_P|\sim
 {|y_\Phi\lambda_A\lambda_3^3\mu_4^2|
 \over M_{11}^2M_3^6M_4^4}.
 \laq{n11matching}
\end{equation}
The coefficient has mass dimension $-10$, as required.  None of the
auxiliary fields needs a vacuum expectation value.  If their masses are
above $T_p$, they are absent from both $g_*$ and $g_{*s}$.

\endgroup
\clearpage
\bibliography{ref
}

\begin{thebibliography}{97}%
\makeatletter
\providecommand \@ifxundefined [1]{%
 \@ifx{#1\undefined}
}%
\providecommand \@ifnum [1]{%
 \ifnum #1\expandafter \@firstoftwo
 \else \expandafter \@secondoftwo
 \fi
}%
\providecommand \@ifx [1]{%
 \ifx #1\expandafter \@firstoftwo
 \else \expandafter \@secondoftwo
 \fi
}%
\providecommand \natexlab [1]{#1}%
\providecommand \enquote  [1]{``#1''}%
\providecommand \bibnamefont  [1]{#1}%
\providecommand \bibfnamefont [1]{#1}%
\providecommand \citenamefont [1]{#1}%
\providecommand \href@noop [0]{\@secondoftwo}%
\providecommand \href [0]{\begingroup \@sanitize@url \@href}%
\providecommand \@href[1]{\@@startlink{#1}\@@href}%
\providecommand \@@href[1]{\endgroup#1\@@endlink}%
\providecommand \@sanitize@url [0]{\catcode `\\12\catcode `\$12\catcode
  `\&12\catcode `\#12\catcode `\^12\catcode `\_12\catcode `\%12\relax}%
\providecommand \@@startlink[1]{}%
\providecommand \@@endlink[0]{}%
\providecommand \url  [0]{\begingroup\@sanitize@url \@url }%
\providecommand \@url [1]{\endgroup\@href {#1}{\urlprefix }}%
\providecommand \urlprefix  [0]{URL }%
\providecommand \Eprint [0]{\href }%
\providecommand \doibase [0]{http://dx.doi.org/}%
\providecommand \selectlanguage [0]{\@gobble}%
\providecommand \bibinfo  [0]{\@secondoftwo}%
\providecommand \bibfield  [0]{\@secondoftwo}%
\providecommand \translation [1]{[#1]}%
\providecommand \BibitemOpen [0]{}%
\providecommand \bibitemStop [0]{}%
\providecommand \bibitemNoStop [0]{.\EOS\space}%
\providecommand \EOS [0]{\spacefactor3000\relax}%
\providecommand \BibitemShut  [1]{\csname bibitem#1\endcsname}%
\let\auto@bib@innerbib\@empty
\bibitem [{\citenamefont {Aghanim}\ \emph {et~al.}(2020)\citenamefont {Aghanim}
  \emph {et~al.}}]{Planck:2018vyg}%
  \BibitemOpen
  \bibfield  {author} {\bibinfo {author} {\bibfnamefont {N.}~\bibnamefont
  {Aghanim}} \emph {et~al.} (\bibinfo {collaboration} {Planck}),\ }\href
  {\doibase 10.1051/0004-6361/201833910} {\bibfield  {journal} {\bibinfo
  {journal} {Astron. Astrophys.}\ }\textbf {\bibinfo {volume} {641}},\ \bibinfo
  {pages} {A6} (\bibinfo {year} {2020})},\ \bibinfo {note} {[Erratum:
  Astron.Astrophys. 652, C4 (2021)]},\ \Eprint
  {http://arxiv.org/abs/1807.06209} {arXiv:1807.06209 [astro-ph.CO]}
  \BibitemShut {NoStop}%
\bibitem [{\citenamefont {Nussinov}(1985)}]{Nussinov:1985xr}%
  \BibitemOpen
  \bibfield  {author} {\bibinfo {author} {\bibfnamefont {S.}~\bibnamefont
  {Nussinov}},\ }\href {\doibase 10.1016/0370-2693(85)90689-6} {\bibfield
  {journal} {\bibinfo  {journal} {Phys. Lett. B}\ }\textbf {\bibinfo {volume}
  {165}},\ \bibinfo {pages} {55} (\bibinfo {year} {1985})}\BibitemShut
  {NoStop}%
\bibitem [{\citenamefont {Kaplan}(1992)}]{Kaplan:1991ah}%
  \BibitemOpen
  \bibfield  {author} {\bibinfo {author} {\bibfnamefont {D.~B.}\ \bibnamefont
  {Kaplan}},\ }\href {\doibase 10.1103/PhysRevLett.68.741} {\bibfield
  {journal} {\bibinfo  {journal} {Phys. Rev. Lett.}\ }\textbf {\bibinfo
  {volume} {68}},\ \bibinfo {pages} {741} (\bibinfo {year} {1992})}\BibitemShut
  {NoStop}%
\bibitem [{\citenamefont {Kaplan}\ \emph {et~al.}(2009)\citenamefont {Kaplan},
  \citenamefont {Luty},\ and\ \citenamefont {Zurek}}]{Kaplan:2009ag}%
  \BibitemOpen
  \bibfield  {author} {\bibinfo {author} {\bibfnamefont {D.~E.}\ \bibnamefont
  {Kaplan}}, \bibinfo {author} {\bibfnamefont {M.~A.}\ \bibnamefont {Luty}}, \
  and\ \bibinfo {author} {\bibfnamefont {K.~M.}\ \bibnamefont {Zurek}},\ }\href
  {\doibase 10.1103/PhysRevD.79.115016} {\bibfield  {journal} {\bibinfo
  {journal} {Phys. Rev. D}\ }\textbf {\bibinfo {volume} {79}},\ \bibinfo
  {pages} {115016} (\bibinfo {year} {2009})},\ \Eprint
  {http://arxiv.org/abs/0901.4117} {arXiv:0901.4117 [hep-ph]} \BibitemShut
  {NoStop}%
\bibitem [{\citenamefont {Zurek}(2014)}]{Zurek:2013wia}%
  \BibitemOpen
  \bibfield  {author} {\bibinfo {author} {\bibfnamefont {K.~M.}\ \bibnamefont
  {Zurek}},\ }\href {\doibase 10.1016/j.physrep.2013.12.001} {\bibfield
  {journal} {\bibinfo  {journal} {Phys. Rept.}\ }\textbf {\bibinfo {volume}
  {537}},\ \bibinfo {pages} {91} (\bibinfo {year} {2014})},\ \Eprint
  {http://arxiv.org/abs/1308.0338} {arXiv:1308.0338 [hep-ph]} \BibitemShut
  {NoStop}%
\bibitem [{\citenamefont {Kitano}\ and\ \citenamefont
  {Low}(2005)}]{Kitano:2004sv}%
  \BibitemOpen
  \bibfield  {author} {\bibinfo {author} {\bibfnamefont {R.}~\bibnamefont
  {Kitano}}\ and\ \bibinfo {author} {\bibfnamefont {I.}~\bibnamefont {Low}},\
  }\href {\doibase 10.1103/PhysRevD.71.023510} {\bibfield  {journal} {\bibinfo
  {journal} {Phys. Rev. D}\ }\textbf {\bibinfo {volume} {71}},\ \bibinfo
  {pages} {023510} (\bibinfo {year} {2005})},\ \Eprint
  {http://arxiv.org/abs/hep-ph/0411133} {arXiv:hep-ph/0411133} \BibitemShut
  {NoStop}%
\bibitem [{\citenamefont {Fujii}\ and\ \citenamefont
  {Yanagida}(2002)}]{Fujii:2002aj}%
  \BibitemOpen
  \bibfield  {author} {\bibinfo {author} {\bibfnamefont {M.}~\bibnamefont
  {Fujii}}\ and\ \bibinfo {author} {\bibfnamefont {T.}~\bibnamefont
  {Yanagida}},\ }\href {\doibase 10.1016/S0370-2693(02)02341-9} {\bibfield
  {journal} {\bibinfo  {journal} {Phys. Lett. B}\ }\textbf {\bibinfo {volume}
  {542}},\ \bibinfo {pages} {80} (\bibinfo {year} {2002})},\ \Eprint
  {http://arxiv.org/abs/hep-ph/0206066} {arXiv:hep-ph/0206066} \BibitemShut
  {NoStop}%
\bibitem [{\citenamefont {Roszkowski}\ and\ \citenamefont
  {Seto}(2007)}]{Roszkowski:2006kw}%
  \BibitemOpen
  \bibfield  {author} {\bibinfo {author} {\bibfnamefont {L.}~\bibnamefont
  {Roszkowski}}\ and\ \bibinfo {author} {\bibfnamefont {O.}~\bibnamefont
  {Seto}},\ }\href {\doibase 10.1103/PhysRevLett.98.161304} {\bibfield
  {journal} {\bibinfo  {journal} {Phys. Rev. Lett.}\ }\textbf {\bibinfo
  {volume} {98}},\ \bibinfo {pages} {161304} (\bibinfo {year} {2007})},\
  \Eprint {http://arxiv.org/abs/hep-ph/0608013} {arXiv:hep-ph/0608013}
  \BibitemShut {NoStop}%
\bibitem [{\citenamefont {Hook}(2014)}]{Hook:2014mla}%
  \BibitemOpen
  \bibfield  {author} {\bibinfo {author} {\bibfnamefont {A.}~\bibnamefont
  {Hook}},\ }\href {\doibase 10.1103/PhysRevD.90.083535} {\bibfield  {journal}
  {\bibinfo  {journal} {Phys. Rev. D}\ }\textbf {\bibinfo {volume} {90}},\
  \bibinfo {pages} {083535} (\bibinfo {year} {2014})},\ \Eprint
  {http://arxiv.org/abs/1404.0113} {arXiv:1404.0113 [hep-ph]} \BibitemShut
  {NoStop}%
\bibitem [{\citenamefont {Foot}\ and\ \citenamefont
  {Volkas}(2003)}]{Foot:2003jt}%
  \BibitemOpen
  \bibfield  {author} {\bibinfo {author} {\bibfnamefont {R.}~\bibnamefont
  {Foot}}\ and\ \bibinfo {author} {\bibfnamefont {R.~R.}\ \bibnamefont
  {Volkas}},\ }\href {\doibase 10.1103/PhysRevD.68.021304} {\bibfield
  {journal} {\bibinfo  {journal} {Phys. Rev. D}\ }\textbf {\bibinfo {volume}
  {68}},\ \bibinfo {pages} {021304} (\bibinfo {year} {2003})},\ \Eprint
  {http://arxiv.org/abs/hep-ph/0304261} {arXiv:hep-ph/0304261} \BibitemShut
  {NoStop}%
\bibitem [{\citenamefont {Newstead}\ and\ \citenamefont
  {TerBeek}(2014)}]{Newstead:2014jva}%
  \BibitemOpen
  \bibfield  {author} {\bibinfo {author} {\bibfnamefont {J.~L.}\ \bibnamefont
  {Newstead}}\ and\ \bibinfo {author} {\bibfnamefont {R.~H.}\ \bibnamefont
  {TerBeek}},\ }\href {\doibase 10.1103/PhysRevD.90.074008} {\bibfield
  {journal} {\bibinfo  {journal} {Phys. Rev. D}\ }\textbf {\bibinfo {volume}
  {90}},\ \bibinfo {pages} {074008} (\bibinfo {year} {2014})},\ \Eprint
  {http://arxiv.org/abs/1405.7427} {arXiv:1405.7427 [hep-ph]} \BibitemShut
  {NoStop}%
\bibitem [{\citenamefont {Farina}(2015)}]{Farina:2015uea}%
  \BibitemOpen
  \bibfield  {author} {\bibinfo {author} {\bibfnamefont {M.}~\bibnamefont
  {Farina}},\ }\href {\doibase 10.1088/1475-7516/2015/11/017} {\bibfield
  {journal} {\bibinfo  {journal} {JCAP}\ }\textbf {\bibinfo {volume} {11}},\
  \bibinfo {pages} {017} (\bibinfo {year} {2015})},\ \Eprint
  {http://arxiv.org/abs/1506.03520} {arXiv:1506.03520 [hep-ph]} \BibitemShut
  {NoStop}%
\bibitem [{\citenamefont {Lonsdale}\ and\ \citenamefont
  {Volkas}(2018)}]{Lonsdale:2018xwd}%
  \BibitemOpen
  \bibfield  {author} {\bibinfo {author} {\bibfnamefont {S.~J.}\ \bibnamefont
  {Lonsdale}}\ and\ \bibinfo {author} {\bibfnamefont {R.~R.}\ \bibnamefont
  {Volkas}},\ }\href {\doibase 10.1103/PhysRevD.97.103510} {\bibfield
  {journal} {\bibinfo  {journal} {Phys. Rev. D}\ }\textbf {\bibinfo {volume}
  {97}},\ \bibinfo {pages} {103510} (\bibinfo {year} {2018})},\ \Eprint
  {http://arxiv.org/abs/1801.05561} {arXiv:1801.05561 [hep-ph]} \BibitemShut
  {NoStop}%
\bibitem [{\citenamefont {Ibe}\ \emph {et~al.}(2019)\citenamefont {Ibe},
  \citenamefont {Kamada}, \citenamefont {Kobayashi}, \citenamefont {Kuwahara},\
  and\ \citenamefont {Nakano}}]{Ibe:2019ena}%
  \BibitemOpen
  \bibfield  {author} {\bibinfo {author} {\bibfnamefont {M.}~\bibnamefont
  {Ibe}}, \bibinfo {author} {\bibfnamefont {A.}~\bibnamefont {Kamada}},
  \bibinfo {author} {\bibfnamefont {S.}~\bibnamefont {Kobayashi}}, \bibinfo
  {author} {\bibfnamefont {T.}~\bibnamefont {Kuwahara}}, \ and\ \bibinfo
  {author} {\bibfnamefont {W.}~\bibnamefont {Nakano}},\ }\href {\doibase
  10.1103/PhysRevD.100.075022} {\bibfield  {journal} {\bibinfo  {journal}
  {Phys. Rev. D}\ }\textbf {\bibinfo {volume} {100}},\ \bibinfo {pages}
  {075022} (\bibinfo {year} {2019})},\ \Eprint
  {http://arxiv.org/abs/1907.03404} {arXiv:1907.03404 [hep-ph]} \BibitemShut
  {NoStop}%
\bibitem [{\citenamefont {Murgui}\ and\ \citenamefont
  {Zurek}(2022)}]{Murgui:2021eqf}%
  \BibitemOpen
  \bibfield  {author} {\bibinfo {author} {\bibfnamefont {C.}~\bibnamefont
  {Murgui}}\ and\ \bibinfo {author} {\bibfnamefont {K.~M.}\ \bibnamefont
  {Zurek}},\ }\href {\doibase 10.1103/PhysRevD.105.095002} {\bibfield
  {journal} {\bibinfo  {journal} {Phys. Rev. D}\ }\textbf {\bibinfo {volume}
  {105}},\ \bibinfo {pages} {095002} (\bibinfo {year} {2022})},\ \Eprint
  {http://arxiv.org/abs/2112.08374} {arXiv:2112.08374 [hep-ph]} \BibitemShut
  {NoStop}%
\bibitem [{\citenamefont {Brzeminski}\ and\ \citenamefont
  {Hook}(2024)}]{Brzeminski:2023wza}%
  \BibitemOpen
  \bibfield  {author} {\bibinfo {author} {\bibfnamefont {D.}~\bibnamefont
  {Brzeminski}}\ and\ \bibinfo {author} {\bibfnamefont {A.}~\bibnamefont
  {Hook}},\ }\href {\doibase 10.1103/PhysRevLett.132.201001} {\bibfield
  {journal} {\bibinfo  {journal} {Phys. Rev. Lett.}\ }\textbf {\bibinfo
  {volume} {132}},\ \bibinfo {pages} {201001} (\bibinfo {year} {2024})},\
  \Eprint {http://arxiv.org/abs/2310.07777} {arXiv:2310.07777 [hep-ph]}
  \BibitemShut {NoStop}%
\bibitem [{\citenamefont {Banerjee}\ \emph {et~al.}(2026)\citenamefont
  {Banerjee}, \citenamefont {Brzeminski},\ and\ \citenamefont
  {Hook}}]{Banerjee:2024xhn}%
  \BibitemOpen
  \bibfield  {author} {\bibinfo {author} {\bibfnamefont {A.}~\bibnamefont
  {Banerjee}}, \bibinfo {author} {\bibfnamefont {D.}~\bibnamefont
  {Brzeminski}}, \ and\ \bibinfo {author} {\bibfnamefont {A.}~\bibnamefont
  {Hook}},\ }\href {\doibase 10.1103/98dc-kpx9} {\bibfield  {journal} {\bibinfo
   {journal} {Phys. Rev. D}\ }\textbf {\bibinfo {volume} {113}},\ \bibinfo
  {pages} {115017} (\bibinfo {year} {2026})},\ \Eprint
  {http://arxiv.org/abs/2410.22412} {arXiv:2410.22412 [hep-ph]} \BibitemShut
  {NoStop}%
\bibitem [{\citenamefont {Davis}\ \emph {et~al.}(1985)\citenamefont {Davis},
  \citenamefont {Efstathiou}, \citenamefont {Frenk},\ and\ \citenamefont
  {White}}]{Davis:1985rj}%
  \BibitemOpen
  \bibfield  {author} {\bibinfo {author} {\bibfnamefont {M.}~\bibnamefont
  {Davis}}, \bibinfo {author} {\bibfnamefont {G.}~\bibnamefont {Efstathiou}},
  \bibinfo {author} {\bibfnamefont {C.~S.}\ \bibnamefont {Frenk}}, \ and\
  \bibinfo {author} {\bibfnamefont {S.~D.~M.}\ \bibnamefont {White}},\ }\href
  {\doibase 10.1086/163168} {\bibfield  {journal} {\bibinfo  {journal}
  {Astrophys. J.}\ }\textbf {\bibinfo {volume} {292}},\ \bibinfo {pages} {371}
  (\bibinfo {year} {1985})}\BibitemShut {NoStop}%
\bibitem [{\citenamefont {Yin}(2023)}]{Yin:2023jjj}%
  \BibitemOpen
  \bibfield  {author} {\bibinfo {author} {\bibfnamefont {W.}~\bibnamefont
  {Yin}},\ }\href {\doibase 10.1007/JHEP05(2023)180} {\bibfield  {journal}
  {\bibinfo  {journal} {JHEP}\ }\textbf {\bibinfo {volume} {05}},\ \bibinfo
  {pages} {180} (\bibinfo {year} {2023})},\ \Eprint
  {http://arxiv.org/abs/2301.08735} {arXiv:2301.08735 [hep-ph]} \BibitemShut
  {NoStop}%
\bibitem [{\citenamefont {Sakurai}\ and\ \citenamefont
  {Yin}(2025)}]{Sakurai:2024apm}%
  \BibitemOpen
  \bibfield  {author} {\bibinfo {author} {\bibfnamefont {K.}~\bibnamefont
  {Sakurai}}\ and\ \bibinfo {author} {\bibfnamefont {W.}~\bibnamefont {Yin}},\
  }\href {\doibase 10.1007/JHEP03(2025)202} {\bibfield  {journal} {\bibinfo
  {journal} {JHEP}\ }\textbf {\bibinfo {volume} {03}},\ \bibinfo {pages} {202}
  (\bibinfo {year} {2025})},\ \Eprint {http://arxiv.org/abs/2410.18968}
  {arXiv:2410.18968 [hep-ph]} \BibitemShut {NoStop}%
\bibitem [{\citenamefont {Moroi}\ and\ \citenamefont
  {Yin}(2021{\natexlab{a}})}]{Moroi:2020has}%
  \BibitemOpen
  \bibfield  {author} {\bibinfo {author} {\bibfnamefont {T.}~\bibnamefont
  {Moroi}}\ and\ \bibinfo {author} {\bibfnamefont {W.}~\bibnamefont {Yin}},\
  }\href {\doibase 10.1007/JHEP03(2021)301} {\bibfield  {journal} {\bibinfo
  {journal} {JHEP}\ }\textbf {\bibinfo {volume} {03}},\ \bibinfo {pages} {301}
  (\bibinfo {year} {2021}{\natexlab{a}})},\ \Eprint
  {http://arxiv.org/abs/2011.09475} {arXiv:2011.09475 [hep-ph]} \BibitemShut
  {NoStop}%
\bibitem [{\citenamefont {Moroi}\ and\ \citenamefont
  {Yin}(2021{\natexlab{b}})}]{Moroi:2020bkq}%
  \BibitemOpen
  \bibfield  {author} {\bibinfo {author} {\bibfnamefont {T.}~\bibnamefont
  {Moroi}}\ and\ \bibinfo {author} {\bibfnamefont {W.}~\bibnamefont {Yin}},\
  }\href {\doibase 10.1007/JHEP03(2021)296} {\bibfield  {journal} {\bibinfo
  {journal} {JHEP}\ }\textbf {\bibinfo {volume} {03}},\ \bibinfo {pages} {296}
  (\bibinfo {year} {2021}{\natexlab{b}})},\ \Eprint
  {http://arxiv.org/abs/2011.12285} {arXiv:2011.12285 [hep-ph]} \BibitemShut
  {NoStop}%
\bibitem [{\citenamefont {Patwardhan}\ \emph {et~al.}(2015)\citenamefont
  {Patwardhan}, \citenamefont {Fuller}, \citenamefont {Kishimoto},\ and\
  \citenamefont {Kusenko}}]{Patwardhan:2015kga}%
  \BibitemOpen
  \bibfield  {author} {\bibinfo {author} {\bibfnamefont {A.~V.}\ \bibnamefont
  {Patwardhan}}, \bibinfo {author} {\bibfnamefont {G.~M.}\ \bibnamefont
  {Fuller}}, \bibinfo {author} {\bibfnamefont {C.~T.}\ \bibnamefont
  {Kishimoto}}, \ and\ \bibinfo {author} {\bibfnamefont {A.}~\bibnamefont
  {Kusenko}},\ }\href {\doibase 10.1103/PhysRevD.92.103509} {\bibfield
  {journal} {\bibinfo  {journal} {Phys. Rev. D}\ }\textbf {\bibinfo {volume}
  {92}},\ \bibinfo {pages} {103509} (\bibinfo {year} {2015})},\ \Eprint
  {http://arxiv.org/abs/1507.01977} {arXiv:1507.01977 [astro-ph.CO]}
  \BibitemShut {NoStop}%
\bibitem [{\citenamefont {Khlebnikov}\ and\ \citenamefont
  {Shaposhnikov}(1988)}]{Khlebnikov:1988sr}%
  \BibitemOpen
  \bibfield  {author} {\bibinfo {author} {\bibfnamefont {S.~Y.}\ \bibnamefont
  {Khlebnikov}}\ and\ \bibinfo {author} {\bibfnamefont {M.~E.}\ \bibnamefont
  {Shaposhnikov}},\ }\href {\doibase 10.1016/0550-3213(88)90133-2} {\bibfield
  {journal} {\bibinfo  {journal} {Nucl. Phys. B}\ }\textbf {\bibinfo {volume}
  {308}},\ \bibinfo {pages} {885} (\bibinfo {year} {1988})}\BibitemShut
  {NoStop}%
\bibitem [{\citenamefont {Harvey}\ and\ \citenamefont
  {Turner}(1990)}]{Harvey:1990qw}%
  \BibitemOpen
  \bibfield  {author} {\bibinfo {author} {\bibfnamefont {J.~A.}\ \bibnamefont
  {Harvey}}\ and\ \bibinfo {author} {\bibfnamefont {M.~S.}\ \bibnamefont
  {Turner}},\ }\href {\doibase 10.1103/PhysRevD.42.3344} {\bibfield  {journal}
  {\bibinfo  {journal} {Phys. Rev. D}\ }\textbf {\bibinfo {volume} {42}},\
  \bibinfo {pages} {3344} (\bibinfo {year} {1990})}\BibitemShut {NoStop}%
\bibitem [{\citenamefont {Burnier}\ \emph {et~al.}(2006)\citenamefont
  {Burnier}, \citenamefont {Laine},\ and\ \citenamefont
  {Shaposhnikov}}]{Burnier:2005hp}%
  \BibitemOpen
  \bibfield  {author} {\bibinfo {author} {\bibfnamefont {Y.}~\bibnamefont
  {Burnier}}, \bibinfo {author} {\bibfnamefont {M.}~\bibnamefont {Laine}}, \
  and\ \bibinfo {author} {\bibfnamefont {M.}~\bibnamefont {Shaposhnikov}},\
  }\href {\doibase 10.1088/1475-7516/2006/02/007} {\bibfield  {journal}
  {\bibinfo  {journal} {JCAP}\ }\textbf {\bibinfo {volume} {02}},\ \bibinfo
  {pages} {007} (\bibinfo {year} {2006})},\ \Eprint
  {http://arxiv.org/abs/hep-ph/0511246} {arXiv:hep-ph/0511246} \BibitemShut
  {NoStop}%
\bibitem [{\citenamefont {Cohen}\ and\ \citenamefont
  {Kaplan}(1987)}]{Cohen:1987vi}%
  \BibitemOpen
  \bibfield  {author} {\bibinfo {author} {\bibfnamefont {A.~G.}\ \bibnamefont
  {Cohen}}\ and\ \bibinfo {author} {\bibfnamefont {D.~B.}\ \bibnamefont
  {Kaplan}},\ }\href {\doibase 10.1016/0370-2693(87)91369-4} {\bibfield
  {journal} {\bibinfo  {journal} {Phys. Lett. B}\ }\textbf {\bibinfo {volume}
  {199}},\ \bibinfo {pages} {251} (\bibinfo {year} {1987})}\BibitemShut
  {NoStop}%
\bibitem [{\citenamefont {Vanvlasselaer}\ and\ \citenamefont
  {Yin}(2026)}]{2604.20762}%
  \BibitemOpen
  \bibfield  {author} {\bibinfo {author} {\bibfnamefont {M.}~\bibnamefont
  {Vanvlasselaer}}\ and\ \bibinfo {author} {\bibfnamefont {W.}~\bibnamefont
  {Yin}},\ }\href@noop {} {\  (\bibinfo {year} {2026})},\ \Eprint
  {http://arxiv.org/abs/2604.20762} {arXiv:2604.20762 [hep-ph]} \BibitemShut
  {NoStop}%
\bibitem [{\citenamefont {Affleck}\ and\ \citenamefont
  {Dine}(1985)}]{Affleck:1984fy}%
  \BibitemOpen
  \bibfield  {author} {\bibinfo {author} {\bibfnamefont {I.}~\bibnamefont
  {Affleck}}\ and\ \bibinfo {author} {\bibfnamefont {M.}~\bibnamefont {Dine}},\
  }\href {\doibase 10.1016/0550-3213(85)90021-5} {\bibfield  {journal}
  {\bibinfo  {journal} {Nucl. Phys. B}\ }\textbf {\bibinfo {volume} {249}},\
  \bibinfo {pages} {361} (\bibinfo {year} {1985})}\BibitemShut {NoStop}%
\bibitem [{\citenamefont {Cohen}\ and\ \citenamefont
  {Kaplan}(1988)}]{Cohen:1988kt}%
  \BibitemOpen
  \bibfield  {author} {\bibinfo {author} {\bibfnamefont {A.~G.}\ \bibnamefont
  {Cohen}}\ and\ \bibinfo {author} {\bibfnamefont {D.~B.}\ \bibnamefont
  {Kaplan}},\ }\href {\doibase 10.1016/0550-3213(88)90134-4} {\bibfield
  {journal} {\bibinfo  {journal} {Nucl. Phys. B}\ }\textbf {\bibinfo {volume}
  {308}},\ \bibinfo {pages} {913} (\bibinfo {year} {1988})}\BibitemShut
  {NoStop}%
\bibitem [{\citenamefont {Davoudiasl}\ \emph {et~al.}(2004)\citenamefont
  {Davoudiasl}, \citenamefont {Kitano}, \citenamefont {Kribs}, \citenamefont
  {Murayama},\ and\ \citenamefont {Steinhardt}}]{Davoudiasl:2004gf}%
  \BibitemOpen
  \bibfield  {author} {\bibinfo {author} {\bibfnamefont {H.}~\bibnamefont
  {Davoudiasl}}, \bibinfo {author} {\bibfnamefont {R.}~\bibnamefont {Kitano}},
  \bibinfo {author} {\bibfnamefont {G.~D.}\ \bibnamefont {Kribs}}, \bibinfo
  {author} {\bibfnamefont {H.}~\bibnamefont {Murayama}}, \ and\ \bibinfo
  {author} {\bibfnamefont {P.~J.}\ \bibnamefont {Steinhardt}},\ }\href
  {\doibase 10.1103/PhysRevLett.93.201301} {\bibfield  {journal} {\bibinfo
  {journal} {Phys. Rev. Lett.}\ }\textbf {\bibinfo {volume} {93}},\ \bibinfo
  {pages} {201301} (\bibinfo {year} {2004})},\ \Eprint
  {http://arxiv.org/abs/hep-ph/0403019} {arXiv:hep-ph/0403019} \BibitemShut
  {NoStop}%
\bibitem [{\citenamefont {Li}\ \emph {et~al.}(2002)\citenamefont {Li},
  \citenamefont {Wang}, \citenamefont {Feng},\ and\ \citenamefont
  {Zhang}}]{Li:2001st}%
  \BibitemOpen
  \bibfield  {author} {\bibinfo {author} {\bibfnamefont {M.-z.}\ \bibnamefont
  {Li}}, \bibinfo {author} {\bibfnamefont {X.-l.}\ \bibnamefont {Wang}},
  \bibinfo {author} {\bibfnamefont {B.}~\bibnamefont {Feng}}, \ and\ \bibinfo
  {author} {\bibfnamefont {X.-m.}\ \bibnamefont {Zhang}},\ }\href {\doibase
  10.1103/PhysRevD.65.103511} {\bibfield  {journal} {\bibinfo  {journal} {Phys.
  Rev. D}\ }\textbf {\bibinfo {volume} {65}},\ \bibinfo {pages} {103511}
  (\bibinfo {year} {2002})},\ \Eprint {http://arxiv.org/abs/hep-ph/0112069}
  {arXiv:hep-ph/0112069} \BibitemShut {NoStop}%
\bibitem [{\citenamefont {Kusenko}\ \emph
  {et~al.}(2015{\natexlab{a}})\citenamefont {Kusenko}, \citenamefont {Pearce},\
  and\ \citenamefont {Yang}}]{Kusenko:2014lra}%
  \BibitemOpen
  \bibfield  {author} {\bibinfo {author} {\bibfnamefont {A.}~\bibnamefont
  {Kusenko}}, \bibinfo {author} {\bibfnamefont {L.}~\bibnamefont {Pearce}}, \
  and\ \bibinfo {author} {\bibfnamefont {L.}~\bibnamefont {Yang}},\ }\href
  {\doibase 10.1103/PhysRevLett.114.061302} {\bibfield  {journal} {\bibinfo
  {journal} {Phys. Rev. Lett.}\ }\textbf {\bibinfo {volume} {114}},\ \bibinfo
  {pages} {061302} (\bibinfo {year} {2015}{\natexlab{a}})},\ \Eprint
  {http://arxiv.org/abs/1410.0722} {arXiv:1410.0722 [hep-ph]} \BibitemShut
  {NoStop}%
\bibitem [{\citenamefont {Kusenko}\ \emph
  {et~al.}(2015{\natexlab{b}})\citenamefont {Kusenko}, \citenamefont
  {Schmitz},\ and\ \citenamefont {Yanagida}}]{Kusenko:2014uta}%
  \BibitemOpen
  \bibfield  {author} {\bibinfo {author} {\bibfnamefont {A.}~\bibnamefont
  {Kusenko}}, \bibinfo {author} {\bibfnamefont {K.}~\bibnamefont {Schmitz}}, \
  and\ \bibinfo {author} {\bibfnamefont {T.~T.}\ \bibnamefont {Yanagida}},\
  }\href {\doibase 10.1103/PhysRevLett.115.011302} {\bibfield  {journal}
  {\bibinfo  {journal} {Phys. Rev. Lett.}\ }\textbf {\bibinfo {volume} {115}},\
  \bibinfo {pages} {011302} (\bibinfo {year} {2015}{\natexlab{b}})},\ \Eprint
  {http://arxiv.org/abs/1412.2043} {arXiv:1412.2043 [hep-ph]} \BibitemShut
  {NoStop}%
\bibitem [{\citenamefont {Ibe}\ and\ \citenamefont
  {Kaneta}(2015)}]{Ibe:2015nfa}%
  \BibitemOpen
  \bibfield  {author} {\bibinfo {author} {\bibfnamefont {M.}~\bibnamefont
  {Ibe}}\ and\ \bibinfo {author} {\bibfnamefont {K.}~\bibnamefont {Kaneta}},\
  }\href {\doibase 10.1103/PhysRevD.92.035019} {\bibfield  {journal} {\bibinfo
  {journal} {Phys. Rev. D}\ }\textbf {\bibinfo {volume} {92}},\ \bibinfo
  {pages} {035019} (\bibinfo {year} {2015})},\ \Eprint
  {http://arxiv.org/abs/1504.04125} {arXiv:1504.04125 [hep-ph]} \BibitemShut
  {NoStop}%
\bibitem [{\citenamefont {Fukugita}\ and\ \citenamefont
  {Yanagida}(1986)}]{Fukugita:1986hr}%
  \BibitemOpen
  \bibfield  {author} {\bibinfo {author} {\bibfnamefont {M.}~\bibnamefont
  {Fukugita}}\ and\ \bibinfo {author} {\bibfnamefont {T.}~\bibnamefont
  {Yanagida}},\ }\href {\doibase 10.1016/0370-2693(86)91126-3} {\bibfield
  {journal} {\bibinfo  {journal} {Phys. Lett. B}\ }\textbf {\bibinfo {volume}
  {174}},\ \bibinfo {pages} {45} (\bibinfo {year} {1986})}\BibitemShut
  {NoStop}%
\bibitem [{\citenamefont {Takahashi}\ and\ \citenamefont
  {Yamaguchi}(2004)}]{Takahashi:2003db}%
  \BibitemOpen
  \bibfield  {author} {\bibinfo {author} {\bibfnamefont {F.}~\bibnamefont
  {Takahashi}}\ and\ \bibinfo {author} {\bibfnamefont {M.}~\bibnamefont
  {Yamaguchi}},\ }\href {\doibase 10.1103/PhysRevD.69.083506} {\bibfield
  {journal} {\bibinfo  {journal} {Phys. Rev. D}\ }\textbf {\bibinfo {volume}
  {69}},\ \bibinfo {pages} {083506} (\bibinfo {year} {2004})},\ \Eprint
  {http://arxiv.org/abs/hep-ph/0308173} {arXiv:hep-ph/0308173} \BibitemShut
  {NoStop}%
\bibitem [{\citenamefont {Chiba}\ \emph {et~al.}(2004)\citenamefont {Chiba},
  \citenamefont {Takahashi},\ and\ \citenamefont {Yamaguchi}}]{Chiba:2003vp}%
  \BibitemOpen
  \bibfield  {author} {\bibinfo {author} {\bibfnamefont {T.}~\bibnamefont
  {Chiba}}, \bibinfo {author} {\bibfnamefont {F.}~\bibnamefont {Takahashi}}, \
  and\ \bibinfo {author} {\bibfnamefont {M.}~\bibnamefont {Yamaguchi}},\ }\href
  {\doibase 10.1103/PhysRevLett.92.011301} {\bibfield  {journal} {\bibinfo
  {journal} {Phys. Rev. Lett.}\ }\textbf {\bibinfo {volume} {92}},\ \bibinfo
  {pages} {011301} (\bibinfo {year} {2004})},\ \bibinfo {note} {[Erratum:
  Phys.Rev.Lett. 114, 209901 (2015)]},\ \Eprint
  {http://arxiv.org/abs/hep-ph/0304102} {arXiv:hep-ph/0304102} \BibitemShut
  {NoStop}%
\bibitem [{\citenamefont {Bond}\ \emph {et~al.}(1980)\citenamefont {Bond},
  \citenamefont {Efstathiou},\ and\ \citenamefont {Silk}}]{Bond:1980ha}%
  \BibitemOpen
  \bibfield  {author} {\bibinfo {author} {\bibfnamefont {J.~R.}\ \bibnamefont
  {Bond}}, \bibinfo {author} {\bibfnamefont {G.}~\bibnamefont {Efstathiou}}, \
  and\ \bibinfo {author} {\bibfnamefont {J.}~\bibnamefont {Silk}},\ }\href
  {\doibase 10.1103/PhysRevLett.45.1980} {\bibfield  {journal} {\bibinfo
  {journal} {Phys. Rev. Lett.}\ }\textbf {\bibinfo {volume} {45}},\ \bibinfo
  {pages} {1980} (\bibinfo {year} {1980})}\BibitemShut {NoStop}%
\bibitem [{\citenamefont {Scherrer}\ and\ \citenamefont
  {Turner}(1985)}]{Scherrer:1984fd}%
  \BibitemOpen
  \bibfield  {author} {\bibinfo {author} {\bibfnamefont {R.~J.}\ \bibnamefont
  {Scherrer}}\ and\ \bibinfo {author} {\bibfnamefont {M.~S.}\ \bibnamefont
  {Turner}},\ }\href {\doibase 10.1103/PhysRevD.31.681} {\bibfield  {journal}
  {\bibinfo  {journal} {Phys. Rev. D}\ }\textbf {\bibinfo {volume} {31}},\
  \bibinfo {pages} {681} (\bibinfo {year} {1985})}\BibitemShut {NoStop}%
\bibitem [{\citenamefont {Bode}\ \emph {et~al.}(2001)\citenamefont {Bode},
  \citenamefont {Ostriker},\ and\ \citenamefont {Turok}}]{Bode:2000gq}%
  \BibitemOpen
  \bibfield  {author} {\bibinfo {author} {\bibfnamefont {P.}~\bibnamefont
  {Bode}}, \bibinfo {author} {\bibfnamefont {J.~P.}\ \bibnamefont {Ostriker}},
  \ and\ \bibinfo {author} {\bibfnamefont {N.}~\bibnamefont {Turok}},\ }\href
  {\doibase 10.1086/321541} {\bibfield  {journal} {\bibinfo  {journal}
  {Astrophys. J.}\ }\textbf {\bibinfo {volume} {556}},\ \bibinfo {pages} {93}
  (\bibinfo {year} {2001})},\ \Eprint {http://arxiv.org/abs/astro-ph/0010389}
  {arXiv:astro-ph/0010389} \BibitemShut {NoStop}%
\bibitem [{\citenamefont {de~Salas}\ \emph {et~al.}(2015)\citenamefont
  {de~Salas}, \citenamefont {Lattanzi}, \citenamefont {Mangano}, \citenamefont
  {Miele}, \citenamefont {Pastor},\ and\ \citenamefont
  {Pisanti}}]{deSalas:2015glj}%
  \BibitemOpen
  \bibfield  {author} {\bibinfo {author} {\bibfnamefont {P.~F.}\ \bibnamefont
  {de~Salas}}, \bibinfo {author} {\bibfnamefont {M.}~\bibnamefont {Lattanzi}},
  \bibinfo {author} {\bibfnamefont {G.}~\bibnamefont {Mangano}}, \bibinfo
  {author} {\bibfnamefont {G.}~\bibnamefont {Miele}}, \bibinfo {author}
  {\bibfnamefont {S.}~\bibnamefont {Pastor}}, \ and\ \bibinfo {author}
  {\bibfnamefont {O.}~\bibnamefont {Pisanti}},\ }\href {\doibase
  10.1103/PhysRevD.92.123534} {\bibfield  {journal} {\bibinfo  {journal} {Phys.
  Rev. D}\ }\textbf {\bibinfo {volume} {92}},\ \bibinfo {pages} {123534}
  (\bibinfo {year} {2015})},\ \Eprint {http://arxiv.org/abs/1511.00672}
  {arXiv:1511.00672 [astro-ph.CO]} \BibitemShut {NoStop}%
\bibitem [{\citenamefont {Domcke}\ \emph {et~al.}(2020)\citenamefont {Domcke},
  \citenamefont {Ema}, \citenamefont {Mukaida},\ and\ \citenamefont
  {Yamada}}]{Domcke:2020kcp}%
  \BibitemOpen
  \bibfield  {author} {\bibinfo {author} {\bibfnamefont {V.}~\bibnamefont
  {Domcke}}, \bibinfo {author} {\bibfnamefont {Y.}~\bibnamefont {Ema}},
  \bibinfo {author} {\bibfnamefont {K.}~\bibnamefont {Mukaida}}, \ and\
  \bibinfo {author} {\bibfnamefont {M.}~\bibnamefont {Yamada}},\ }\href
  {\doibase 10.1007/JHEP08(2020)096} {\bibfield  {journal} {\bibinfo  {journal}
  {JHEP}\ }\textbf {\bibinfo {volume} {08}},\ \bibinfo {pages} {096} (\bibinfo
  {year} {2020})},\ \Eprint {http://arxiv.org/abs/2006.03148} {arXiv:2006.03148
  [hep-ph]} \BibitemShut {NoStop}%
\bibitem [{\citenamefont {D'Onofrio}\ \emph {et~al.}(2014)\citenamefont
  {D'Onofrio}, \citenamefont {Rummukainen},\ and\ \citenamefont
  {Tranberg}}]{DOnofrio:2014rug}%
  \BibitemOpen
  \bibfield  {author} {\bibinfo {author} {\bibfnamefont {M.}~\bibnamefont
  {D'Onofrio}}, \bibinfo {author} {\bibfnamefont {K.}~\bibnamefont
  {Rummukainen}}, \ and\ \bibinfo {author} {\bibfnamefont {A.}~\bibnamefont
  {Tranberg}},\ }\href {\doibase 10.1103/PhysRevLett.113.141602} {\bibfield
  {journal} {\bibinfo  {journal} {Phys. Rev. Lett.}\ }\textbf {\bibinfo
  {volume} {113}},\ \bibinfo {pages} {141602} (\bibinfo {year} {2014})},\
  \Eprint {http://arxiv.org/abs/1404.3565} {arXiv:1404.3565 [hep-ph]}
  \BibitemShut {NoStop}%
\bibitem [{\citenamefont {Jeong}\ \emph {et~al.}(2019)\citenamefont {Jeong},
  \citenamefont {Jung},\ and\ \citenamefont {Shin}}]{Jeong:2018ucz}%
  \BibitemOpen
  \bibfield  {author} {\bibinfo {author} {\bibfnamefont {K.~S.}\ \bibnamefont
  {Jeong}}, \bibinfo {author} {\bibfnamefont {T.~H.}\ \bibnamefont {Jung}}, \
  and\ \bibinfo {author} {\bibfnamefont {C.~S.}\ \bibnamefont {Shin}},\ }\href
  {\doibase 10.1016/j.physletb.2019.01.036} {\bibfield  {journal} {\bibinfo
  {journal} {Phys. Lett. B}\ }\textbf {\bibinfo {volume} {790}},\ \bibinfo
  {pages} {326} (\bibinfo {year} {2019})},\ \Eprint
  {http://arxiv.org/abs/1806.02591} {arXiv:1806.02591 [hep-ph]} \BibitemShut
  {NoStop}%
\bibitem [{\citenamefont {Preskill}\ \emph {et~al.}(1983)\citenamefont
  {Preskill}, \citenamefont {Wise},\ and\ \citenamefont
  {Wilczek}}]{Preskill:1982cy}%
  \BibitemOpen
  \bibfield  {author} {\bibinfo {author} {\bibfnamefont {J.}~\bibnamefont
  {Preskill}}, \bibinfo {author} {\bibfnamefont {M.~B.}\ \bibnamefont {Wise}},
  \ and\ \bibinfo {author} {\bibfnamefont {F.}~\bibnamefont {Wilczek}},\ }\href
  {\doibase 10.1016/0370-2693(83)90637-8} {\bibfield  {journal} {\bibinfo
  {journal} {Phys. Lett. B}\ }\textbf {\bibinfo {volume} {120}},\ \bibinfo
  {pages} {127} (\bibinfo {year} {1983})}\BibitemShut {NoStop}%
\bibitem [{\citenamefont {Abbott}\ and\ \citenamefont
  {Sikivie}(1983)}]{Abbott:1982af}%
  \BibitemOpen
  \bibfield  {author} {\bibinfo {author} {\bibfnamefont {L.~F.}\ \bibnamefont
  {Abbott}}\ and\ \bibinfo {author} {\bibfnamefont {P.}~\bibnamefont
  {Sikivie}},\ }\href {\doibase 10.1016/0370-2693(83)90638-X} {\bibfield
  {journal} {\bibinfo  {journal} {Phys. Lett. B}\ }\textbf {\bibinfo {volume}
  {120}},\ \bibinfo {pages} {133} (\bibinfo {year} {1983})}\BibitemShut
  {NoStop}%
\bibitem [{\citenamefont {Dine}\ and\ \citenamefont
  {Fischler}(1983)}]{Dine:1982ah}%
  \BibitemOpen
  \bibfield  {author} {\bibinfo {author} {\bibfnamefont {M.}~\bibnamefont
  {Dine}}\ and\ \bibinfo {author} {\bibfnamefont {W.}~\bibnamefont
  {Fischler}},\ }\href {\doibase 10.1016/0370-2693(83)90639-1} {\bibfield
  {journal} {\bibinfo  {journal} {Phys. Lett. B}\ }\textbf {\bibinfo {volume}
  {120}},\ \bibinfo {pages} {137} (\bibinfo {year} {1983})}\BibitemShut
  {NoStop}%
\bibitem [{\citenamefont {Minkowski}(1977)}]{Minkowski:1977sc}%
  \BibitemOpen
  \bibfield  {author} {\bibinfo {author} {\bibfnamefont {P.}~\bibnamefont
  {Minkowski}},\ }\href {\doibase 10.1016/0370-2693(77)90435-X} {\bibfield
  {journal} {\bibinfo  {journal} {Phys. Lett. B}\ }\textbf {\bibinfo {volume}
  {67}},\ \bibinfo {pages} {421} (\bibinfo {year} {1977})}\BibitemShut
  {NoStop}%
\bibitem [{\citenamefont {Yanagida}(1979)}]{Yanagida:1979as}%
  \BibitemOpen
  \bibfield  {author} {\bibinfo {author} {\bibfnamefont {T.}~\bibnamefont
  {Yanagida}},\ }\href@noop {} {\bibfield  {journal} {\bibinfo  {journal}
  {Conf. Proc. C}\ }\textbf {\bibinfo {volume} {7902131}},\ \bibinfo {pages}
  {95} (\bibinfo {year} {1979})}\BibitemShut {NoStop}%
\bibitem [{\citenamefont {Gell-Mann}\ \emph {et~al.}(1979)\citenamefont
  {Gell-Mann}, \citenamefont {Ramond},\ and\ \citenamefont
  {Slansky}}]{Gell-Mann:1979vob}%
  \BibitemOpen
  \bibfield  {author} {\bibinfo {author} {\bibfnamefont {M.}~\bibnamefont
  {Gell-Mann}}, \bibinfo {author} {\bibfnamefont {P.}~\bibnamefont {Ramond}}, \
  and\ \bibinfo {author} {\bibfnamefont {R.}~\bibnamefont {Slansky}},\
  }\href@noop {} {\bibfield  {journal} {\bibinfo  {journal} {Conf. Proc. C}\
  }\textbf {\bibinfo {volume} {790927}},\ \bibinfo {pages} {315} (\bibinfo
  {year} {1979})},\ \Eprint {http://arxiv.org/abs/1306.4669} {arXiv:1306.4669
  [hep-th]} \BibitemShut {NoStop}%
\bibitem [{\citenamefont {Mohapatra}\ and\ \citenamefont
  {Senjanovic}(1980)}]{Mohapatra:1979ia}%
  \BibitemOpen
  \bibfield  {author} {\bibinfo {author} {\bibfnamefont {R.~N.}\ \bibnamefont
  {Mohapatra}}\ and\ \bibinfo {author} {\bibfnamefont {G.}~\bibnamefont
  {Senjanovic}},\ }\href {\doibase 10.1103/PhysRevLett.44.912} {\bibfield
  {journal} {\bibinfo  {journal} {Phys. Rev. Lett.}\ }\textbf {\bibinfo
  {volume} {44}},\ \bibinfo {pages} {912} (\bibinfo {year} {1980})}\BibitemShut
  {NoStop}%
\bibitem [{\citenamefont {Akhmedov}\ \emph {et~al.}(1998)\citenamefont
  {Akhmedov}, \citenamefont {Rubakov},\ and\ \citenamefont
  {Smirnov}}]{Akhmedov:1998qx}%
  \BibitemOpen
  \bibfield  {author} {\bibinfo {author} {\bibfnamefont {E.~K.}\ \bibnamefont
  {Akhmedov}}, \bibinfo {author} {\bibfnamefont {V.~A.}\ \bibnamefont
  {Rubakov}}, \ and\ \bibinfo {author} {\bibfnamefont {A.~Y.}\ \bibnamefont
  {Smirnov}},\ }\href {\doibase 10.1103/PhysRevLett.81.1359} {\bibfield
  {journal} {\bibinfo  {journal} {Phys. Rev. Lett.}\ }\textbf {\bibinfo
  {volume} {81}},\ \bibinfo {pages} {1359} (\bibinfo {year} {1998})},\ \Eprint
  {http://arxiv.org/abs/hep-ph/9803255} {arXiv:hep-ph/9803255} \BibitemShut
  {NoStop}%
\bibitem [{\citenamefont {Boyarsky}\ \emph {et~al.}(2021)\citenamefont
  {Boyarsky}, \citenamefont {Ovchynnikov}, \citenamefont {Ruchayskiy},\ and\
  \citenamefont {Syvolap}}]{Boyarsky:2020dzc}%
  \BibitemOpen
  \bibfield  {author} {\bibinfo {author} {\bibfnamefont {A.}~\bibnamefont
  {Boyarsky}}, \bibinfo {author} {\bibfnamefont {M.}~\bibnamefont
  {Ovchynnikov}}, \bibinfo {author} {\bibfnamefont {O.}~\bibnamefont
  {Ruchayskiy}}, \ and\ \bibinfo {author} {\bibfnamefont {V.}~\bibnamefont
  {Syvolap}},\ }\href {\doibase 10.1103/PhysRevD.104.023517} {\bibfield
  {journal} {\bibinfo  {journal} {Phys. Rev. D}\ }\textbf {\bibinfo {volume}
  {104}},\ \bibinfo {pages} {023517} (\bibinfo {year} {2021})},\ \Eprint
  {http://arxiv.org/abs/2008.00749} {arXiv:2008.00749 [hep-ph]} \BibitemShut
  {NoStop}%
\bibitem [{\citenamefont {Coleman}\ and\ \citenamefont
  {Weinberg}(1973)}]{Coleman:1973jx}%
  \BibitemOpen
  \bibfield  {author} {\bibinfo {author} {\bibfnamefont {S.~R.}\ \bibnamefont
  {Coleman}}\ and\ \bibinfo {author} {\bibfnamefont {E.~J.}\ \bibnamefont
  {Weinberg}},\ }\href {\doibase 10.1103/PhysRevD.7.1888} {\bibfield  {journal}
  {\bibinfo  {journal} {Phys. Rev. D}\ }\textbf {\bibinfo {volume} {7}},\
  \bibinfo {pages} {1888} (\bibinfo {year} {1973})}\BibitemShut {NoStop}%
\bibitem [{\citenamefont {Abreu}\ \emph {et~al.}(1997)\citenamefont {Abreu}
  \emph {et~al.}}]{DELPHI:1996qcc}%
  \BibitemOpen
  \bibfield  {author} {\bibinfo {author} {\bibfnamefont {P.}~\bibnamefont
  {Abreu}} \emph {et~al.} (\bibinfo {collaboration} {DELPHI}),\ }\href
  {\doibase 10.1007/s002880050370} {\bibfield  {journal} {\bibinfo  {journal}
  {Z. Phys. C}\ }\textbf {\bibinfo {volume} {74}},\ \bibinfo {pages} {57}
  (\bibinfo {year} {1997})},\ \bibinfo {note} {[Erratum: Z.Phys.C 75, 580
  (1997)]}\BibitemShut {NoStop}%
\bibitem [{\citenamefont {Hayrapetyan}\ \emph {et~al.}(2024)\citenamefont
  {Hayrapetyan} \emph {et~al.}}]{CMS:2023jqi}%
  \BibitemOpen
  \bibfield  {author} {\bibinfo {author} {\bibfnamefont {A.}~\bibnamefont
  {Hayrapetyan}} \emph {et~al.} (\bibinfo {collaboration} {CMS}),\ }\href
  {\doibase 10.1007/JHEP03(2024)105} {\bibfield  {journal} {\bibinfo  {journal}
  {JHEP}\ }\textbf {\bibinfo {volume} {03}},\ \bibinfo {pages} {105} (\bibinfo
  {year} {2024})},\ \Eprint {http://arxiv.org/abs/2312.07484} {arXiv:2312.07484
  [hep-ex]} \BibitemShut {NoStop}%
\bibitem [{\citenamefont {Sakurai}\ and\ \citenamefont
  {Yin}(2022)}]{Sakurai:2021ipp}%
  \BibitemOpen
  \bibfield  {author} {\bibinfo {author} {\bibfnamefont {K.}~\bibnamefont
  {Sakurai}}\ and\ \bibinfo {author} {\bibfnamefont {W.}~\bibnamefont {Yin}},\
  }\href {\doibase 10.1007/JHEP04(2022)113} {\bibfield  {journal} {\bibinfo
  {journal} {JHEP}\ }\textbf {\bibinfo {volume} {04}},\ \bibinfo {pages} {113}
  (\bibinfo {year} {2022})},\ \Eprint {http://arxiv.org/abs/2111.03653}
  {arXiv:2111.03653 [hep-ph]} \BibitemShut {NoStop}%
\bibitem [{\citenamefont {Haghighat}\ \emph {et~al.}(2023)\citenamefont
  {Haghighat}, \citenamefont {Mohammadi~Najafabadi}, \citenamefont {Sakurai},\
  and\ \citenamefont {Yin}}]{Haghighat:2022qyh}%
  \BibitemOpen
  \bibfield  {author} {\bibinfo {author} {\bibfnamefont {G.}~\bibnamefont
  {Haghighat}}, \bibinfo {author} {\bibfnamefont {M.}~\bibnamefont
  {Mohammadi~Najafabadi}}, \bibinfo {author} {\bibfnamefont {K.}~\bibnamefont
  {Sakurai}}, \ and\ \bibinfo {author} {\bibfnamefont {W.}~\bibnamefont
  {Yin}},\ }\href {\doibase 10.1103/PhysRevD.107.035033} {\bibfield  {journal}
  {\bibinfo  {journal} {Phys. Rev. D}\ }\textbf {\bibinfo {volume} {107}},\
  \bibinfo {pages} {035033} (\bibinfo {year} {2023})},\ \Eprint
  {http://arxiv.org/abs/2209.07565} {arXiv:2209.07565 [hep-ph]} \BibitemShut
  {NoStop}%
\bibitem [{\citenamefont {Yin}(2025{\natexlab{a}})}]{Yin:2024txg}%
  \BibitemOpen
  \bibfield  {author} {\bibinfo {author} {\bibfnamefont {W.}~\bibnamefont
  {Yin}},\ }\href {\doibase 10.1007/JHEP10(2025)177} {\bibfield  {journal}
  {\bibinfo  {journal} {JHEP}\ }\textbf {\bibinfo {volume} {10}},\ \bibinfo
  {pages} {177} (\bibinfo {year} {2025}{\natexlab{a}})},\ \Eprint
  {http://arxiv.org/abs/2412.17802} {arXiv:2412.17802 [hep-ph]} \BibitemShut
  {NoStop}%
\bibitem [{\citenamefont {Yin}(2025{\natexlab{b}})}]{Yin:2024pri}%
  \BibitemOpen
  \bibfield  {author} {\bibinfo {author} {\bibfnamefont {W.}~\bibnamefont
  {Yin}},\ }\href {\doibase 10.1093/ptep/ptaf053} {\bibfield  {journal}
  {\bibinfo  {journal} {PTEP}\ }\textbf {\bibinfo {volume} {2025}},\ \bibinfo
  {pages} {053B02} (\bibinfo {year} {2025}{\natexlab{b}})},\ \Eprint
  {http://arxiv.org/abs/2412.19798} {arXiv:2412.19798 [hep-ph]} \BibitemShut
  {NoStop}%
\bibitem [{\citenamefont {Choi}\ and\ \citenamefont {Im}(2016)}]{Choi:2015fiu}%
  \BibitemOpen
  \bibfield  {author} {\bibinfo {author} {\bibfnamefont {K.}~\bibnamefont
  {Choi}}\ and\ \bibinfo {author} {\bibfnamefont {S.~H.}\ \bibnamefont {Im}},\
  }\href {\doibase 10.1007/JHEP01(2016)149} {\bibfield  {journal} {\bibinfo
  {journal} {JHEP}\ }\textbf {\bibinfo {volume} {01}},\ \bibinfo {pages} {149}
  (\bibinfo {year} {2016})},\ \Eprint {http://arxiv.org/abs/1511.00132}
  {arXiv:1511.00132 [hep-ph]} \BibitemShut {NoStop}%
\bibitem [{\citenamefont {Kaplan}\ and\ \citenamefont
  {Rattazzi}(2016)}]{Kaplan:2015fuy}%
  \BibitemOpen
  \bibfield  {author} {\bibinfo {author} {\bibfnamefont {D.~E.}\ \bibnamefont
  {Kaplan}}\ and\ \bibinfo {author} {\bibfnamefont {R.}~\bibnamefont
  {Rattazzi}},\ }\href {\doibase 10.1103/PhysRevD.93.085007} {\bibfield
  {journal} {\bibinfo  {journal} {Phys. Rev. D}\ }\textbf {\bibinfo {volume}
  {93}},\ \bibinfo {pages} {085007} (\bibinfo {year} {2016})},\ \Eprint
  {http://arxiv.org/abs/1511.01827} {arXiv:1511.01827 [hep-ph]} \BibitemShut
  {NoStop}%
\bibitem [{\citenamefont {Higaki}\ \emph {et~al.}(2016)\citenamefont {Higaki},
  \citenamefont {Jeong}, \citenamefont {Kitajima},\ and\ \citenamefont
  {Takahashi}}]{Higaki:2015jag}%
  \BibitemOpen
  \bibfield  {author} {\bibinfo {author} {\bibfnamefont {T.}~\bibnamefont
  {Higaki}}, \bibinfo {author} {\bibfnamefont {K.~S.}\ \bibnamefont {Jeong}},
  \bibinfo {author} {\bibfnamefont {N.}~\bibnamefont {Kitajima}}, \ and\
  \bibinfo {author} {\bibfnamefont {F.}~\bibnamefont {Takahashi}},\ }\href
  {\doibase 10.1016/j.physletb.2016.01.055} {\bibfield  {journal} {\bibinfo
  {journal} {Phys. Lett. B}\ }\textbf {\bibinfo {volume} {755}},\ \bibinfo
  {pages} {13} (\bibinfo {year} {2016})},\ \Eprint
  {http://arxiv.org/abs/1512.05295} {arXiv:1512.05295 [hep-ph]} \BibitemShut
  {NoStop}%
\bibitem [{\citenamefont {Giudice}\ and\ \citenamefont
  {McCullough}(2017)}]{Giudice:2016yja}%
  \BibitemOpen
  \bibfield  {author} {\bibinfo {author} {\bibfnamefont {G.~F.}\ \bibnamefont
  {Giudice}}\ and\ \bibinfo {author} {\bibfnamefont {M.}~\bibnamefont
  {McCullough}},\ }\href {\doibase 10.1007/JHEP02(2017)036} {\bibfield
  {journal} {\bibinfo  {journal} {JHEP}\ }\textbf {\bibinfo {volume} {02}},\
  \bibinfo {pages} {036} (\bibinfo {year} {2017})},\ \Eprint
  {http://arxiv.org/abs/1610.07962} {arXiv:1610.07962 [hep-ph]} \BibitemShut
  {NoStop}%
\bibitem [{\citenamefont {Ir{{s}}i{{c}}}\ \emph {et~al.}(2017)\citenamefont
  {Ir{{s}}i{{c}}} \emph {et~al.}}]{Irsic:2017ixq}%
  \BibitemOpen
  \bibfield  {author} {\bibinfo {author} {\bibfnamefont {V.}~\bibnamefont
  {Ir{{s}}i{{c}}}} \emph {et~al.},\ }\href {\doibase
  10.1103/PhysRevD.96.023522} {\bibfield  {journal} {\bibinfo  {journal} {Phys.
  Rev. D}\ }\textbf {\bibinfo {volume} {96}},\ \bibinfo {pages} {023522}
  (\bibinfo {year} {2017})},\ \Eprint {http://arxiv.org/abs/1702.01764}
  {arXiv:1702.01764 [astro-ph.CO]} \BibitemShut {NoStop}%
\bibitem [{\citenamefont {Bessho}\ \emph {et~al.}(2022)\citenamefont {Bessho},
  \citenamefont {Ikeda},\ and\ \citenamefont {Yin}}]{Bessho:2022yyu}%
  \BibitemOpen
  \bibfield  {author} {\bibinfo {author} {\bibfnamefont {T.}~\bibnamefont
  {Bessho}}, \bibinfo {author} {\bibfnamefont {Y.}~\bibnamefont {Ikeda}}, \
  and\ \bibinfo {author} {\bibfnamefont {W.}~\bibnamefont {Yin}},\ }\href
  {\doibase 10.1103/PhysRevD.106.095025} {\bibfield  {journal} {\bibinfo
  {journal} {Phys. Rev. D}\ }\textbf {\bibinfo {volume} {106}},\ \bibinfo
  {pages} {095025} (\bibinfo {year} {2022})},\ \Eprint
  {http://arxiv.org/abs/2208.05975} {arXiv:2208.05975 [hep-ph]} \BibitemShut
  {NoStop}%
\bibitem [{\citenamefont {Homma}\ \emph {et~al.}(2023)\citenamefont {Homma},
  \citenamefont {Ishibashi}, \citenamefont {Kirita},\ and\ \citenamefont
  {Hasada}}]{Homma:2022ktv}%
  \BibitemOpen
  \bibfield  {author} {\bibinfo {author} {\bibfnamefont {K.}~\bibnamefont
  {Homma}}, \bibinfo {author} {\bibfnamefont {F.}~\bibnamefont {Ishibashi}},
  \bibinfo {author} {\bibfnamefont {Y.}~\bibnamefont {Kirita}}, \ and\ \bibinfo
  {author} {\bibfnamefont {T.}~\bibnamefont {Hasada}},\ }\href {\doibase
  10.3390/universe9010020} {\bibfield  {journal} {\bibinfo  {journal}
  {Universe}\ }\textbf {\bibinfo {volume} {9}},\ \bibinfo {pages} {20}
  (\bibinfo {year} {2023})},\ \Eprint {http://arxiv.org/abs/2212.13012}
  {arXiv:2212.13012 [hep-ph]} \BibitemShut {NoStop}%
\bibitem [{\citenamefont {Yin}\ and\ \citenamefont
  {Yoshida}(2025)}]{Yin:2024rjb}%
  \BibitemOpen
  \bibfield  {author} {\bibinfo {author} {\bibfnamefont {W.}~\bibnamefont
  {Yin}}\ and\ \bibinfo {author} {\bibfnamefont {J.}~\bibnamefont {Yoshida}},\
  }\href {\doibase 10.1103/PhysRevD.111.036020} {\bibfield  {journal} {\bibinfo
   {journal} {Phys. Rev. D}\ }\textbf {\bibinfo {volume} {111}},\ \bibinfo
  {pages} {036020} (\bibinfo {year} {2025})},\ \Eprint
  {http://arxiv.org/abs/2408.17451} {arXiv:2408.17451 [hep-ph]} \BibitemShut
  {NoStop}%
\bibitem [{\citenamefont {Yin}(2026)}]{Yin:2025bui}%
  \BibitemOpen
  \bibfield  {author} {\bibinfo {author} {\bibfnamefont {W.}~\bibnamefont
  {Yin}},\ }\href {\doibase 10.1103/snnn-wqxg} {\bibfield  {journal} {\bibinfo
  {journal} {Phys. Rev. Lett.}\ }\textbf {\bibinfo {volume} {136}},\ \bibinfo
  {pages} {131803} (\bibinfo {year} {2026})},\ \Eprint
  {http://arxiv.org/abs/2508.14885} {arXiv:2508.14885 [hep-ph]} \BibitemShut
  {NoStop}%
\bibitem [{\citenamefont {Jaeckel}\ and\ \citenamefont
  {Ringwald}(2010)}]{Jaeckel:2010ni}%
  \BibitemOpen
  \bibfield  {author} {\bibinfo {author} {\bibfnamefont {J.}~\bibnamefont
  {Jaeckel}}\ and\ \bibinfo {author} {\bibfnamefont {A.}~\bibnamefont
  {Ringwald}},\ }\href {\doibase 10.1146/annurev.nucl.012809.104433} {\bibfield
   {journal} {\bibinfo  {journal} {Ann. Rev. Nucl. Part. Sci.}\ }\textbf
  {\bibinfo {volume} {60}},\ \bibinfo {pages} {405} (\bibinfo {year} {2010})},\
  \Eprint {http://arxiv.org/abs/1002.0329} {arXiv:1002.0329 [hep-ph]}
  \BibitemShut {NoStop}%
\bibitem [{\citenamefont {Ringwald}(2012)}]{Ringwald:2012hr}%
  \BibitemOpen
  \bibfield  {author} {\bibinfo {author} {\bibfnamefont {A.}~\bibnamefont
  {Ringwald}},\ }\href {\doibase 10.1016/j.dark.2012.10.008} {\bibfield
  {journal} {\bibinfo  {journal} {Phys. Dark Univ.}\ }\textbf {\bibinfo
  {volume} {1}},\ \bibinfo {pages} {116} (\bibinfo {year} {2012})},\ \Eprint
  {http://arxiv.org/abs/1210.5081} {arXiv:1210.5081 [hep-ph]} \BibitemShut
  {NoStop}%
\bibitem [{\citenamefont {Arias}\ \emph {et~al.}(2012)\citenamefont {Arias},
  \citenamefont {Cadamuro}, \citenamefont {Goodsell}, \citenamefont {Jaeckel},
  \citenamefont {Redondo},\ and\ \citenamefont {Ringwald}}]{Arias:2012az}%
  \BibitemOpen
  \bibfield  {author} {\bibinfo {author} {\bibfnamefont {P.}~\bibnamefont
  {Arias}}, \bibinfo {author} {\bibfnamefont {D.}~\bibnamefont {Cadamuro}},
  \bibinfo {author} {\bibfnamefont {M.}~\bibnamefont {Goodsell}}, \bibinfo
  {author} {\bibfnamefont {J.}~\bibnamefont {Jaeckel}}, \bibinfo {author}
  {\bibfnamefont {J.}~\bibnamefont {Redondo}}, \ and\ \bibinfo {author}
  {\bibfnamefont {A.}~\bibnamefont {Ringwald}},\ }\href {\doibase
  10.1088/1475-7516/2012/06/013} {\bibfield  {journal} {\bibinfo  {journal}
  {JCAP}\ }\textbf {\bibinfo {volume} {06}},\ \bibinfo {pages} {013} (\bibinfo
  {year} {2012})},\ \Eprint {http://arxiv.org/abs/1201.5902} {arXiv:1201.5902
  [hep-ph]} \BibitemShut {NoStop}%
\bibitem [{\citenamefont {Graham}\ \emph {et~al.}(2015)\citenamefont {Graham},
  \citenamefont {Irastorza}, \citenamefont {Lamoreaux}, \citenamefont
  {Lindner},\ and\ \citenamefont {van Bibber}}]{Graham:2015ouw}%
  \BibitemOpen
  \bibfield  {author} {\bibinfo {author} {\bibfnamefont {P.~W.}\ \bibnamefont
  {Graham}}, \bibinfo {author} {\bibfnamefont {I.~G.}\ \bibnamefont
  {Irastorza}}, \bibinfo {author} {\bibfnamefont {S.~K.}\ \bibnamefont
  {Lamoreaux}}, \bibinfo {author} {\bibfnamefont {A.}~\bibnamefont {Lindner}},
  \ and\ \bibinfo {author} {\bibfnamefont {K.~A.}\ \bibnamefont {van Bibber}},\
  }\href {\doibase 10.1146/annurev-nucl-102014-022120} {\bibfield  {journal}
  {\bibinfo  {journal} {Ann. Rev. Nucl. Part. Sci.}\ }\textbf {\bibinfo
  {volume} {65}},\ \bibinfo {pages} {485} (\bibinfo {year} {2015})},\ \Eprint
  {http://arxiv.org/abs/1602.00039} {arXiv:1602.00039 [hep-ex]} \BibitemShut
  {NoStop}%
\bibitem [{\citenamefont {Marsh}(2016)}]{Marsh:2015xka}%
  \BibitemOpen
  \bibfield  {author} {\bibinfo {author} {\bibfnamefont {D.~J.~E.}\
  \bibnamefont {Marsh}},\ }\href {\doibase 10.1016/j.physrep.2016.06.005}
  {\bibfield  {journal} {\bibinfo  {journal} {Phys. Rept.}\ }\textbf {\bibinfo
  {volume} {643}},\ \bibinfo {pages} {1} (\bibinfo {year} {2016})},\ \Eprint
  {http://arxiv.org/abs/1510.07633} {arXiv:1510.07633 [astro-ph.CO]}
  \BibitemShut {NoStop}%
\bibitem [{\citenamefont {Irastorza}\ and\ \citenamefont
  {Redondo}(2018)}]{Irastorza:2018dyq}%
  \BibitemOpen
  \bibfield  {author} {\bibinfo {author} {\bibfnamefont {I.~G.}\ \bibnamefont
  {Irastorza}}\ and\ \bibinfo {author} {\bibfnamefont {J.}~\bibnamefont
  {Redondo}},\ }\href {\doibase 10.1016/j.ppnp.2018.05.003} {\bibfield
  {journal} {\bibinfo  {journal} {Prog. Part. Nucl. Phys.}\ }\textbf {\bibinfo
  {volume} {102}},\ \bibinfo {pages} {89} (\bibinfo {year} {2018})},\ \Eprint
  {http://arxiv.org/abs/1801.08127} {arXiv:1801.08127 [hep-ph]} \BibitemShut
  {NoStop}%
\bibitem [{\citenamefont {Di~Luzio}\ \emph {et~al.}(2020)\citenamefont
  {Di~Luzio}, \citenamefont {Giannotti}, \citenamefont {Nardi},\ and\
  \citenamefont {Visinelli}}]{DiLuzio:2020wdo}%
  \BibitemOpen
  \bibfield  {author} {\bibinfo {author} {\bibfnamefont {L.}~\bibnamefont
  {Di~Luzio}}, \bibinfo {author} {\bibfnamefont {M.}~\bibnamefont {Giannotti}},
  \bibinfo {author} {\bibfnamefont {E.}~\bibnamefont {Nardi}}, \ and\ \bibinfo
  {author} {\bibfnamefont {L.}~\bibnamefont {Visinelli}},\ }\href {\doibase
  10.1016/j.physrep.2020.06.002} {\bibfield  {journal} {\bibinfo  {journal}
  {Phys. Rept.}\ }\textbf {\bibinfo {volume} {870}},\ \bibinfo {pages} {1}
  (\bibinfo {year} {2020})},\ \Eprint {http://arxiv.org/abs/2003.01100}
  {arXiv:2003.01100 [hep-ph]} \BibitemShut {NoStop}%
\bibitem [{\citenamefont {Albertus}\ \emph {et~al.}(2026)\citenamefont
  {Albertus} \emph {et~al.}}]{Albertus:2026fbe}%
  \BibitemOpen
  \bibfield  {author} {\bibinfo {author} {\bibfnamefont {C.}~\bibnamefont
  {Albertus}} \emph {et~al.},\ }\href@noop {} {\  (\bibinfo {year} {2026})},\
  \Eprint {http://arxiv.org/abs/2602.09089} {arXiv:2602.09089 [hep-ph]}
  \BibitemShut {NoStop}%
\bibitem [{\citenamefont {Arza}\ \emph {et~al.}(2026)\citenamefont {Arza} \emph
  {et~al.}}]{Arza:2026rsl}%
  \BibitemOpen
  \bibfield  {author} {\bibinfo {author} {\bibfnamefont {A.}~\bibnamefont
  {Arza}} \emph {et~al.},\ }\href@noop {} {\  (\bibinfo {year} {2026})},\
  \Eprint {http://arxiv.org/abs/2603.03433} {arXiv:2603.03433 [hep-ph]}
  \BibitemShut {NoStop}%
\bibitem [{\citenamefont {Sigl}\ and\ \citenamefont
  {Raffelt}(1993)}]{Sigl:1993ctk}%
  \BibitemOpen
  \bibfield  {author} {\bibinfo {author} {\bibfnamefont {G.}~\bibnamefont
  {Sigl}}\ and\ \bibinfo {author} {\bibfnamefont {G.}~\bibnamefont {Raffelt}},\
  }\href {\doibase 10.1016/0550-3213(93)90175-O} {\bibfield  {journal}
  {\bibinfo  {journal} {Nucl. Phys. B}\ }\textbf {\bibinfo {volume} {406}},\
  \bibinfo {pages} {423} (\bibinfo {year} {1993})}\BibitemShut {NoStop}%
\bibitem [{\citenamefont {Drewes}\ \emph {et~al.}(2016)\citenamefont {Drewes},
  \citenamefont {Garbrecht}, \citenamefont {Gueter},\ and\ \citenamefont
  {Klaric}}]{Drewes:2016gmt}%
  \BibitemOpen
  \bibfield  {author} {\bibinfo {author} {\bibfnamefont {M.}~\bibnamefont
  {Drewes}}, \bibinfo {author} {\bibfnamefont {B.}~\bibnamefont {Garbrecht}},
  \bibinfo {author} {\bibfnamefont {D.}~\bibnamefont {Gueter}}, \ and\ \bibinfo
  {author} {\bibfnamefont {J.}~\bibnamefont {Klaric}},\ }\href {\doibase
  10.1007/JHEP12(2016)150} {\bibfield  {journal} {\bibinfo  {journal} {JHEP}\
  }\textbf {\bibinfo {volume} {12}},\ \bibinfo {pages} {150} (\bibinfo {year}
  {2016})},\ \Eprint {http://arxiv.org/abs/1606.06690} {arXiv:1606.06690
  [hep-ph]} \BibitemShut {NoStop}%
\bibitem [{\citenamefont {Hamada}\ \emph {et~al.}(2018)\citenamefont {Hamada},
  \citenamefont {Kitano},\ and\ \citenamefont {Yin}}]{Hamada:2018epb}%
  \BibitemOpen
  \bibfield  {author} {\bibinfo {author} {\bibfnamefont {Y.}~\bibnamefont
  {Hamada}}, \bibinfo {author} {\bibfnamefont {R.}~\bibnamefont {Kitano}}, \
  and\ \bibinfo {author} {\bibfnamefont {W.}~\bibnamefont {Yin}},\ }\href
  {\doibase 10.1007/JHEP10(2018)178} {\bibfield  {journal} {\bibinfo  {journal}
  {JHEP}\ }\textbf {\bibinfo {volume} {10}},\ \bibinfo {pages} {178} (\bibinfo
  {year} {2018})},\ \Eprint {http://arxiv.org/abs/1807.06582} {arXiv:1807.06582
  [hep-ph]} \BibitemShut {NoStop}%
\bibitem [{\citenamefont {Ghiglieri}\ and\ \citenamefont
  {Laine}(2017)}]{Ghiglieri:2017gjz}%
  \BibitemOpen
  \bibfield  {author} {\bibinfo {author} {\bibfnamefont {J.}~\bibnamefont
  {Ghiglieri}}\ and\ \bibinfo {author} {\bibfnamefont {M.}~\bibnamefont
  {Laine}},\ }\href {\doibase 10.1007/JHEP05(2017)132} {\bibfield  {journal}
  {\bibinfo  {journal} {JHEP}\ }\textbf {\bibinfo {volume} {05}},\ \bibinfo
  {pages} {132} (\bibinfo {year} {2017})},\ \Eprint
  {http://arxiv.org/abs/1703.06087} {arXiv:1703.06087 [hep-ph]} \BibitemShut
  {NoStop}%
\bibitem [{\citenamefont {Yin}\ \emph {et~al.}(2025)\citenamefont {Yin},
  \citenamefont {Nakagawa}, \citenamefont {Murokoshi},\ and\ \citenamefont
  {Hattori}}]{Yin:2024trc}%
  \BibitemOpen
  \bibfield  {author} {\bibinfo {author} {\bibfnamefont {W.}~\bibnamefont
  {Yin}}, \bibinfo {author} {\bibfnamefont {S.}~\bibnamefont {Nakagawa}},
  \bibinfo {author} {\bibfnamefont {T.}~\bibnamefont {Murokoshi}}, \ and\
  \bibinfo {author} {\bibfnamefont {M.}~\bibnamefont {Hattori}},\ }\href
  {\doibase 10.1088/1475-7516/2025/02/063} {\bibfield  {journal} {\bibinfo
  {journal} {JCAP}\ }\textbf {\bibinfo {volume} {02}},\ \bibinfo {pages} {063}
  (\bibinfo {year} {2025})},\ \Eprint {http://arxiv.org/abs/2405.10303}
  {arXiv:2405.10303 [hep-ph]} \BibitemShut {NoStop}%
\bibitem [{\citenamefont {Ghiglieri}\ and\ \citenamefont
  {Laine}(2016)}]{Ghiglieri:2016xye}%
  \BibitemOpen
  \bibfield  {author} {\bibinfo {author} {\bibfnamefont {J.}~\bibnamefont
  {Ghiglieri}}\ and\ \bibinfo {author} {\bibfnamefont {M.}~\bibnamefont
  {Laine}},\ }\href {\doibase 10.1088/1475-7516/2016/07/015} {\bibfield
  {journal} {\bibinfo  {journal} {JCAP}\ }\textbf {\bibinfo {volume} {07}},\
  \bibinfo {pages} {015} (\bibinfo {year} {2016})},\ \Eprint
  {http://arxiv.org/abs/1605.07720} {arXiv:1605.07720 [hep-ph]} \BibitemShut
  {NoStop}%
\bibitem [{\citenamefont {Eijima}\ and\ \citenamefont
  {Shaposhnikov}(2017)}]{Eijima:2017anv}%
  \BibitemOpen
  \bibfield  {author} {\bibinfo {author} {\bibfnamefont {S.}~\bibnamefont
  {Eijima}}\ and\ \bibinfo {author} {\bibfnamefont {M.}~\bibnamefont
  {Shaposhnikov}},\ }\href {\doibase 10.1016/j.physletb.2017.05.068} {\bibfield
   {journal} {\bibinfo  {journal} {Phys. Lett. B}\ }\textbf {\bibinfo {volume}
  {771}},\ \bibinfo {pages} {288} (\bibinfo {year} {2017})},\ \Eprint
  {http://arxiv.org/abs/1703.06085} {arXiv:1703.06085 [hep-ph]} \BibitemShut
  {NoStop}%
\bibitem [{\citenamefont {Laine}(2022)}]{Laine:2022pgk}%
  \BibitemOpen
  \bibfield  {author} {\bibinfo {author} {\bibfnamefont {M.}~\bibnamefont
  {Laine}},\ }\href {\doibase 10.1016/j.aop.2022.169022} {\bibfield  {journal}
  {\bibinfo  {journal} {Annals Phys.}\ }\textbf {\bibinfo {volume} {444}},\
  \bibinfo {pages} {169022} (\bibinfo {year} {2022})},\ \Eprint
  {http://arxiv.org/abs/2203.05772} {arXiv:2203.05772 [hep-ph]} \BibitemShut
  {NoStop}%
\bibitem [{\citenamefont {Hernandez}\ \emph {et~al.}(2022)\citenamefont
  {Hernandez}, \citenamefont {Lopez-Pavon}, \citenamefont {Rius},\ and\
  \citenamefont {Sandner}}]{Hernandez:2022ivz}%
  \BibitemOpen
  \bibfield  {author} {\bibinfo {author} {\bibfnamefont {P.}~\bibnamefont
  {Hernandez}}, \bibinfo {author} {\bibfnamefont {J.}~\bibnamefont
  {Lopez-Pavon}}, \bibinfo {author} {\bibfnamefont {N.}~\bibnamefont {Rius}}, \
  and\ \bibinfo {author} {\bibfnamefont {S.}~\bibnamefont {Sandner}},\ }\href
  {\doibase 10.1007/JHEP12(2022)012} {\bibfield  {journal} {\bibinfo  {journal}
  {JHEP}\ }\textbf {\bibinfo {volume} {12}},\ \bibinfo {pages} {012} (\bibinfo
  {year} {2022})},\ \Eprint {http://arxiv.org/abs/2207.01651} {arXiv:2207.01651
  [hep-ph]} \BibitemShut {NoStop}%
\bibitem [{\citenamefont {Frampton}\ \emph {et~al.}(2002)\citenamefont
  {Frampton}, \citenamefont {Glashow},\ and\ \citenamefont
  {Yanagida}}]{Frampton:2002qc}%
  \BibitemOpen
  \bibfield  {author} {\bibinfo {author} {\bibfnamefont {P.~H.}\ \bibnamefont
  {Frampton}}, \bibinfo {author} {\bibfnamefont {S.~L.}\ \bibnamefont
  {Glashow}}, \ and\ \bibinfo {author} {\bibfnamefont {T.}~\bibnamefont
  {Yanagida}},\ }\href {\doibase 10.1016/S0370-2693(02)02853-8} {\bibfield
  {journal} {\bibinfo  {journal} {Phys. Lett. B}\ }\textbf {\bibinfo {volume}
  {548}},\ \bibinfo {pages} {119} (\bibinfo {year} {2002})},\ \Eprint
  {http://arxiv.org/abs/hep-ph/0208157} {arXiv:hep-ph/0208157} \BibitemShut
  {NoStop}%
\bibitem [{\citenamefont {Rink}\ \emph {et~al.}(2016)\citenamefont {Rink},
  \citenamefont {Schmitz},\ and\ \citenamefont {Yanagida}}]{Rink:2016knw}%
  \BibitemOpen
  \bibfield  {author} {\bibinfo {author} {\bibfnamefont {T.}~\bibnamefont
  {Rink}}, \bibinfo {author} {\bibfnamefont {K.}~\bibnamefont {Schmitz}}, \
  and\ \bibinfo {author} {\bibfnamefont {T.~T.}\ \bibnamefont {Yanagida}},\
  }\href@noop {} {\  (\bibinfo {year} {2016})},\ \Eprint
  {http://arxiv.org/abs/1612.08878} {arXiv:1612.08878 [hep-ph]} \BibitemShut
  {NoStop}%
\bibitem [{\citenamefont {Cordero-Carri{\'o}n}\ \emph
  {et~al.}(2020)\citenamefont {Cordero-Carri{\'o}n}, \citenamefont {Hirsch},\
  and\ \citenamefont {Vicente}}]{Cordero-Carrion:2019qtu}%
  \BibitemOpen
  \bibfield  {author} {\bibinfo {author} {\bibfnamefont {I.}~\bibnamefont
  {Cordero-Carri{\'o}n}}, \bibinfo {author} {\bibfnamefont {M.}~\bibnamefont
  {Hirsch}}, \ and\ \bibinfo {author} {\bibfnamefont {A.}~\bibnamefont
  {Vicente}},\ }\href {\doibase 10.1103/PhysRevD.101.075032} {\bibfield
  {journal} {\bibinfo  {journal} {Phys. Rev. D}\ }\textbf {\bibinfo {volume}
  {101}},\ \bibinfo {pages} {075032} (\bibinfo {year} {2020})},\ \Eprint
  {http://arxiv.org/abs/1912.08858} {arXiv:1912.08858 [hep-ph]} \BibitemShut
  {NoStop}%
\bibitem [{\citenamefont {Casas}\ and\ \citenamefont
  {Ibarra}(2001)}]{Casas:2001sr}%
  \BibitemOpen
  \bibfield  {author} {\bibinfo {author} {\bibfnamefont {J.~A.}\ \bibnamefont
  {Casas}}\ and\ \bibinfo {author} {\bibfnamefont {A.}~\bibnamefont {Ibarra}},\
  }\href {\doibase 10.1016/S0550-3213(01)00475-8} {\bibfield  {journal}
  {\bibinfo  {journal} {Nucl. Phys. B}\ }\textbf {\bibinfo {volume} {618}},\
  \bibinfo {pages} {171} (\bibinfo {year} {2001})},\ \Eprint
  {http://arxiv.org/abs/hep-ph/0103065} {arXiv:hep-ph/0103065} \BibitemShut
  {NoStop}%
\bibitem [{\citenamefont {Esteban}\ \emph {et~al.}(2024)\citenamefont
  {Esteban}, \citenamefont {Gonzalez-Garcia}, \citenamefont {Maltoni},
  \citenamefont {Martinez-Soler}, \citenamefont {Pinheiro},\ and\ \citenamefont
  {Schwetz}}]{Esteban:2024eli}%
  \BibitemOpen
  \bibfield  {author} {\bibinfo {author} {\bibfnamefont {I.}~\bibnamefont
  {Esteban}}, \bibinfo {author} {\bibfnamefont {M.~C.}\ \bibnamefont
  {Gonzalez-Garcia}}, \bibinfo {author} {\bibfnamefont {M.}~\bibnamefont
  {Maltoni}}, \bibinfo {author} {\bibfnamefont {I.}~\bibnamefont
  {Martinez-Soler}}, \bibinfo {author} {\bibfnamefont {J.~P.}\ \bibnamefont
  {Pinheiro}}, \ and\ \bibinfo {author} {\bibfnamefont {T.}~\bibnamefont
  {Schwetz}},\ }\href {\doibase 10.1007/JHEP12(2024)216} {\bibfield  {journal}
  {\bibinfo  {journal} {JHEP}\ }\textbf {\bibinfo {volume} {12}},\ \bibinfo
  {pages} {216} (\bibinfo {year} {2024})},\ \Eprint
  {http://arxiv.org/abs/2410.05380} {arXiv:2410.05380 [hep-ph]} \BibitemShut
  {NoStop}%
\bibitem [{\citenamefont {Dev}\ and\ \citenamefont
  {Pilaftsis}(2012)}]{Dev:2012sg}%
  \BibitemOpen
  \bibfield  {author} {\bibinfo {author} {\bibfnamefont {P.~S.~B.}\
  \bibnamefont {Dev}}\ and\ \bibinfo {author} {\bibfnamefont {A.}~\bibnamefont
  {Pilaftsis}},\ }\href {\doibase 10.1103/PhysRevD.86.113001} {\bibfield
  {journal} {\bibinfo  {journal} {Phys. Rev. D}\ }\textbf {\bibinfo {volume}
  {86}},\ \bibinfo {pages} {113001} (\bibinfo {year} {2012})},\ \Eprint
  {http://arxiv.org/abs/1209.4051} {arXiv:1209.4051 [hep-ph]} \BibitemShut
  {NoStop}%
\bibitem [{\citenamefont {Lopez-Pavon}\ \emph {et~al.}(2015)\citenamefont
  {Lopez-Pavon}, \citenamefont {Molinaro},\ and\ \citenamefont
  {Petcov}}]{Lopez-Pavon:2015cga}%
  \BibitemOpen
  \bibfield  {author} {\bibinfo {author} {\bibfnamefont {J.}~\bibnamefont
  {Lopez-Pavon}}, \bibinfo {author} {\bibfnamefont {E.}~\bibnamefont
  {Molinaro}}, \ and\ \bibinfo {author} {\bibfnamefont {S.~T.}\ \bibnamefont
  {Petcov}},\ }\href {\doibase 10.1007/JHEP11(2015)030} {\bibfield  {journal}
  {\bibinfo  {journal} {JHEP}\ }\textbf {\bibinfo {volume} {11}},\ \bibinfo
  {pages} {030} (\bibinfo {year} {2015})},\ \Eprint
  {http://arxiv.org/abs/1506.05296} {arXiv:1506.05296 [hep-ph]} \BibitemShut
  {NoStop}%
\bibitem [{\citenamefont {Alvi}\ \emph {et~al.}(2022)\citenamefont {Alvi},
  \citenamefont {Brinckmann}, \citenamefont {Gerbino}, \citenamefont
  {Lattanzi},\ and\ \citenamefont {Pagano}}]{Alvi:2022aam}%
  \BibitemOpen
  \bibfield  {author} {\bibinfo {author} {\bibfnamefont {S.}~\bibnamefont
  {Alvi}}, \bibinfo {author} {\bibfnamefont {T.}~\bibnamefont {Brinckmann}},
  \bibinfo {author} {\bibfnamefont {M.}~\bibnamefont {Gerbino}}, \bibinfo
  {author} {\bibfnamefont {M.}~\bibnamefont {Lattanzi}}, \ and\ \bibinfo
  {author} {\bibfnamefont {L.}~\bibnamefont {Pagano}},\ }\href {\doibase
  10.1088/1475-7516/2022/11/015} {\bibfield  {journal} {\bibinfo  {journal}
  {JCAP}\ }\textbf {\bibinfo {volume} {11}},\ \bibinfo {pages} {015} (\bibinfo
  {year} {2022})},\ \Eprint {http://arxiv.org/abs/2205.05636} {arXiv:2205.05636
  [astro-ph.CO]} \BibitemShut {NoStop}%
\bibitem [{\citenamefont {McKeen}(2019)}]{McKeen:2018xyz}%
  \BibitemOpen
  \bibfield  {author} {\bibinfo {author} {\bibfnamefont {D.}~\bibnamefont
  {McKeen}},\ }\href {\doibase 10.1103/PhysRevD.100.015028} {\bibfield
  {journal} {\bibinfo  {journal} {Phys. Rev. D}\ }\textbf {\bibinfo {volume}
  {100}},\ \bibinfo {pages} {015028} (\bibinfo {year} {2019})},\ \Eprint
  {http://arxiv.org/abs/1812.08178} {arXiv:1812.08178 [hep-ph]} \BibitemShut
  {NoStop}%
\end{thebibliography}%
\end{document}